\documentstyle[preprint,prb,aps]{revtex}

\begin{document}
\tightenlines
\title{\bf Effective interactions and superconductivity in the t-J model
in the large-N limit}

\vspace{3cm}
\author{{R. Zeyher}$^a$ and {A. Greco}$^{a,b}$}

\address{$^a$Max-Planck-Institut\  f\"ur\
Festk\"orperforschung,\\ Heisenbergstr.1, 70569 Stuttgart, Germany \\
$^b$Permanent address: Departamento de F\'{\i}sica, Facultad de
Ciencias Exactas e Ingenier\'{\i}a and IFIR(UNR-CONICET), Av. Pellegrini 250, 
2000-Rosario, Argentina}

\date{\today }

\vspace{3cm}

\maketitle

\begin{abstract}

The feasibility of a perturbation expansion for Green's functions
of the $t-J$ model directly in terms of X-operators is demonstrated 
using the Baym-Kadanoff
functional method. As an application we derive explicit expressions 
for the kernel $\Theta$ of the
linearized equation for the superconducting order parameter
in leading order of a 1/N expansion. The linearized equation is
solved numerically on a square lattice taking 
instantaneous and retarded contributions into account. \par
Classifying the order parameter according to irreducible representations
$\Gamma_i,i=1,...5,$ of the point group $C_{4v}$ of the square lattice
and according to even or odd parity in frequency we find that a resonably 
strong
instability occurs only for even frequency pairing with d-wavelike
$\Gamma_3$ symmetry. The corresponding transition temperature $T_c$ is
$\sim 0.01 |t|$ where $t$ is the nearest-neighbor hopping integral.
The underlying effective interaction consists of an attractive,
instantaneous term and a retarded term due to charge and spin fluctuations.
The latter is weakly attractive at low frequencies below $\sim J/2$,
strongly repulsive up to $\sim |t|$ and attractive towards even higher
energies. $T_c$ increases with decreasing doping $\delta$ until a
d-wavelike bond-order wave instability is encountered near optimal doping
at $\delta_{BO} \sim 0.14$ for $J=0.3$. $T_c$ is essentially linear
in $J$ and rather insensitive to an additional second-nearest neighbor
hopping integral $t'$. A rather striking property of $T_c$ is that it is
hardly affected by the soft mode associated with the bond-order wave
instability or by the Van Hove singularity in the case with second-nearest
neighbor hopping. This unique feature reflects the fact that the solution
of the gap equation involves momenta far away from the Fermi surface
(due to the instantaneous term) and many frequencies (due to the
retarded term) so that singular properties in momentum or frequency
are averaged out very effectively.

\end{abstract}

\newpage 

\section{Introduction}
 Many studies of the $t-J$ model suggest that this model is able to 
describe a large body of the low-energy physics of real high-T$_c$
superconductors \cite{Anderson1,Dagotto1}. 
This is true for many normal state properties
of high-T$_c$ oxides where accurate numerical predictions of the $t-J$
model are available for the comparison with experiment. Whether
the phenomenon of high-T$_c$ superconductivity itself can be explained
within this model is presently not so clear. There are several
calculations yielding instabilities of the normal state with respect 
to d-wave superconductivity  \cite{Dagotto1,Grilli1,Houghton1,Heeb1}
and also reasonably high values for the
transition temperature $T_c$ \cite{Dagotto2,Onufrieva1,Plakida1,Zeyher1}.
These calculations, however, use often
somewhat uncontrolled assumptions, making definite conclusions
difficult. There is also the view \cite{Anderson1} that the large 
observed values
for $T_c$ are not directly related to large mean field
$T_c$'s in isolated $CuO_2$ planes described by the
$t-J$ model. Instead it is argued that pair tunneling 
between planes enhances strongly weak, plane-related superconducting
instabilities producing in this way the phenomen of high-T$_c$ 
superconductivity. \par

In this paper we present a new attempt to calculate $T_c$ for the
$t-J$ model in a well controlled way. Similar as in Refs. 
\cite{Grilli1,Houghton1}
we do not assume the validity of Migdal's theorem or 
approximate the self-energy by the lowest skeleton graphs.
Instead we assume that $1/N$ can be considered as a small parameter
where $N$ is the number of electronic degrees of freedom per site.
$N$ consists of two spin directions times $N/2$ copies of the
local electronic orbital counted by a flavor index. Similar like
in many slave boson calculations \cite{Grilli1,Houghton1,Kotliar1}
the flavor index is introduced in a
somewhat artificial way just to make $N$ a large integer. The
$SU(2)$ spin symmetry of the original model is thus enlarged to
the symplectic group  $Sp(N/2)$. The original constraint of having
no double occupancies of sites is modified to the condition that
at most $N/2$ electrons can occupy the $N$ states at each site.
Compared to Refs. \cite{Grilli1,Houghton1} our treatment exhibits 
two novel features. First, 
the constraint is implemented in a different and more
rigorous way yielding differences in the equation for the superconducting gap
already in the leading order O(1/N). Secondly, 
we have solved the linearized gap equation numerically obtaining
also values for $T_c$. Our conclusions about the occurrence of
superconductivity in the $t-J$ model are thus no longer based only
on Fermi surface averaged, static coupling strengths as in Refs.
\cite{Grilli1,Houghton1,Kotliar1,Greco1}. \par

Regarding constraints their implementation in the X-operator
approach is trivial \cite{Zeyher2}. Mathematically, the constraint 
means that
the sum over diagonal elements of X-operators at a given site
has to be equal to $N/2$. This sum commutes with every X-operator
and thus is a multiple of the identity operator in any irreducible
representation of X-operators. Enforcing the constraint means
therefore just to select the correct subspace of Hilberts space where
the eigenvalue of this sum is equal to $N/2$. In slave boson theory
the X-operators are represented by product of slave operators and
the Hilbert space is enlarged to the Hilbert space of slave particles.
The constraint means now that at each site the number operator of slave 
particles has to be equal $N/2$. This condition can obviously not
hold as an operator identity, i.e., cannot be enforced at any
position of an expectation value of slave operators. However, it should
be enforced at all positions which originally separated X-operators.
Such an enforcement for a fixed $N$ seems to be in conflict
with the Bose condensation of the bosonic slave particles as well as
the independence of fermion and bosonic slaves in the limit $N \rightarrow
\infty$. Using the Dirac method to enforce constraints on the operator
level it has indeed been shown \cite{Guillou1} that the commutator relations
for the slave particles have to be changed and that, for instance,
the fermionc and the bosonic slave operators no longer commute with
each other. In view of this open problems we use in sections II and III
a perturbation expansion directly in X-operators following the work
of Ref. \cite{Ruckenstein1} This procedure will give us also
the opportunity to compare the
O(1/N) expression for the gap equation with that of the slave boson 
approach to see whether the $1/N$ expansions are really the same
in the two cases. The final analytic results for the gap equation
have already been presented in Ref. \cite{Zeyher1}, however, without
derivations. These derivations can be found in sections II and III.   
\par
The second new feature of our work deals with the solution of
the gap equation. Previous work using 1/N expansions concluded
from Fermi surface averaged, static coupling constants on the 
occurrence of superconductivity. Our previous work \cite{Zeyher1,Greco1}
has shown
that superconducting instabilities occur for any symmetry channel,
doping and both for $J = 0$ and $J \neq 0$. The point is more 
whether the corresponding $T_c$ is very small or large enough 
to be relevant. $T_c$ here is understood as the transition temperature
within mean-field theory. This means that the lowering of $T_c$ due to
fluctuations in the superconducting order parameter is not taken
into account. This assumption is justified if one compares with
real, three-dimensional superconductors but does not apply, of course,
to strictly two-dimensional models where fluctuations drive the order 
parameter to zero at any finite temperature.  
In order to determine $T_c$ one has to solve the
gap equation which has several non-BCS features: The kernel of the 
linearized gap equation consists of an instantaneous and a
retarded term and both are characterized by different cutoffs.
Moreover, the presence of the instantaneous term does not allow
to restrict the momenta to the Fermi surface. Consequently, we solved
the gap equation by numerical means. Our method is also suitable to
investigate
the question of odd frequency pairing \cite{Abrahams1} in our model. 
In this case
the instantaneous term drops out but the full frequency dependence of the
kernel must be kept in addition to the momentum dependence along the 
Fermi line. Results for odd frequency
pairing will also be presented in section IV together with the conclusions.

\section{Model and general equations for the electron Green's function}
The Hamiltonian of the $t-J$ model can be written as
\begin{equation}
H = \sum_{{ij} \atop {p=1...N}} {{t_{ij}} \over N} X_i^{p0} X_j^{0p}
+ \sum_{{ij} \atop {p,q=1...N}} {{J_{ij}} \over {4N}} X_i^{pq}
X_j^{qp} 
- \sum_{{ij} \atop {p,q=1...N}} {{J_{ij}} \over {4N}} X_i^{pp}
X_j^{qq}. 
\end{equation}
Let us consider first the case $N=2$. Assuming one orbital per site and
excluding double occupancies of sites there are three states $|{p \atop i}>$
at each atom $i$. $p=0$ denotes the empty state, and $p=1,2$ singly
occupied states with spin up and down. The Hubbard operators $X^{pq}_i$
can be represented as projection operators $X^{pq}_i = {|{p \atop i}>}
{<{q \atop i}|}$ and obey the following commutator and anticommutator
relations
\begin{equation}
[ X_i^{pq}, X_j^{rs}]_{\pm} = \delta_{ij} (\delta_{qr}X_i^{ps}
\pm \delta_{sr} X_i^{rq}),
\end{equation}
and the completeness relation
\begin{equation}
\sum_{p=0}^2 X_i^{pp} = 1.
\end{equation}
The upper (lower) sign in Eq.(2) holds for bosonlike or mixed (fermionlike)
Hubbard operators defined by $p,q>0$ or $p=q=0$ ($p=0,q>0$ or $p>0,q=0$).
For $i=j$ both the upper and lower signs hold in each case.
The first term in Eq.(1) describes
the hopping of particles between the sites $i$ and $j$ with 
matrix elements $t_{ij}$. The second term in Eq.(1) denotes
the Heisenberg interaction between the spin densities at site $i$ and
$j$ with the exchange constants $J_{ij}$. The third term in Eq.(1)
represents the charge-charge interaction of the $t-J$ model.
In the following we 
consider $J_{ij}$ only between nearest neighbors ($J_{ij}=J$)
and $t_{ij}$ between nearest ($t_{ij}=t$) and next nearest
($t_{ij}=t'$) neighbors. We also use always $|t|$ as energy unit. \par

The Hamiltonian in Eq.(1) is a generalization from $N=2$ to an (even) 
arbitrary integer $N$. The orbital index $p$ consists now of the spin 
and a flavor index, the latter enumerating $N/2$ identical orbitals at a
site. The symmetry group of $H$ is the symplectic group $Sp(N/2)$.
For $N>2$ the operators $X$ can no longer be written as projection
operators in some basis. Instead, fermionlike (bosonlike or mixed ones)
operators are assumed to satisfy the anticommutator (commutator)
relations of Eq.(2). Only some of the diagonal operators can assumed
to retain their projection properties, namely, $({X_i^{pp}})^2=X_i^{pp}$
for $p>0$. 
Most other relations characteristic of projection operators
such as $X^{10}_i X^{01}_i=X_i^{11}$ are lost for $N>2$. The completeness
relation Eq.(3) is replaced by the constraint
\begin{equation}
Q_i = \sum_{p=0}^N X_i^{pp} = N/2.
\end{equation}
By explicit construction of the Hilbert space and the action
of the $X's$ on its vectors one can show \cite{Zeyher2}
that the above properties, together
with $H$, specify completely the problem. In particular, $X_i^{00}$
is a non-negative operator. As a result, Eq.(4) means that at most
$N/2$ particles can occupy the $N$ available states at a site. In this way
one may expect that expectation values of observables approach
smoothly the physical case $N=2$ from large $N's$ yielding a basis for
$1/N$ expansions. We also would like to point out that slave boson
treatments of $H$ have many features in common with our approach. However,
the Hilbert spaces are different in the two cases as well as the
enforcement of the constraint Eq.(4). $Q_i$ commutes with all Hubbard
operators and thus is proportional to the identiy in any irreducible
representation of the X-operators. Enforcement of the constraint means
in our case just the selection of the correct subspace of the Hilbert
space where Eq.(4) is satisfied as an operator identity. \par
Following the Baym-Kadanoff procedure \cite{Baym1,Zeyher3,Kulic1}
we define a non-equilibrium 
Matsubara Green's function of two fermionic operators $X(1)$ and $X(2)$ by
\begin{equation}
G(12) = -<TSX(1)X(2)>/<S>,
\end{equation}
\begin{equation}
S = Te^{\int d1 X(1) K(1)}.
\end{equation}
The number $1$ stands for an internal pair index $p,q$ as well as
a site and a (imaginary) time index $i,\tau$. $\int d1$ means $\sum_{p,q,i}
\int_0^\beta d\tau$, where $\beta$ is the inverse temperature.
In Eqs.(5) and (6) $T$ is the time  ordering operator and $K$ an external 
source term which is assumed to couple only to bosonic X-operators.
We also introduce the non-equilibirum expectation value of bosonic
X-operators by
\begin{equation}
L(1) = <TSX(1)>/<S>.
\end{equation}
Using the Hamiltonian Eq.(1) the Heisenberg equation of motion for 
fermionic operators becomes 
\begin{equation}
{\partial \over {\partial \tau_1}} X(1) = \int d2 d3 t(123) X(2)X(3),
\end{equation}
with
$$
t(123) = \delta(\tau_2-\tau_1) \delta(\tau_3-\tau_1)
{(t_{i_1i_3}\delta_{i_1i_2}+J_{i_1i_2}/2\delta_{i_1i_3})  
\over N} \Big[\delta_{q_10}\delta_{q_30}(1-\delta_{p_30})
\delta_{p_20} \delta_{q_2q_1} \delta_{p_3p_1}+
\delta_{p_2p_1}\delta_{q_2p_3}) 
$$
\begin{equation}
-\delta_{p_10} \delta_{p_30}(1-\delta_{q_30})
(\delta_{p_2q_3}
\delta_{q_2q_1} +\delta_{p_2p_1} \delta_{q_20}
\delta_{q_3q_1}) \Big]. 
\end{equation}
Eq.(9) agrees with Eq.(5) of Ref. \cite{Zeyher4} 
if the different sign convention for
the hopping term is taken into account.
Using Eq.(8) and rewriting higher-order correlation functions in terms
of functional derivatives with respect to $K$
the equation of motion for $G$ can be written as
\begin{equation}
\int d2 ( G_0^{-1}(12) -\Sigma^{'}(12))G(21') = Q(11'),
\end{equation}
\begin{equation}
\Sigma^{'}(12) = -\int d3 t(132)L(3) +\int d3 d4 d5 t(134)G(45)\Gamma(52,3),
\end{equation}
\begin{equation}
\Gamma(12,3) = {\delta G^{-1}(12)}/ {\delta K(3)}.
\end{equation}
$G_0$ is the unperturbed Green's function
\begin{equation}
G^{-1}_0(12) = -\delta(1-2) {\partial \over{\partial \tau_2}} -
\delta({\bar 1}-{\bar 2})(K^{00}({\bar 1})\delta_{q_1q_2}-K^{q_1q_2}
({\bar 1})).
\end{equation}
$Q$ is given by 
\begin{equation}
Q(11') = \delta(1-1')(L^{pq'}({\bar 1}) \delta_{qp'}+L^{p'q}({\bar 1})
\delta_{q'p}),
\end{equation}
where the index $1=(pq,i,\tau)$ has been split into $1=(pq,{\bar 1})$
with ${\bar 1}=(i \tau)$. $L$ can be expressed by $G$ so the above
system of equations for $G$, the self-energy $\Sigma^{'}$, and the
vertex $\Gamma$ is closed. \par
For our purposes it is more convenient to use a normalized Green's function 
$g$ which obeys a Dyson equation with the usual $\delta$-function on its
right-hand side. Writing
\begin{equation}
g(11') = \int d2 G(12) Z^{-1}(21'),
\end{equation}
and requesting that Z satisfies the following equation
\begin{equation}
Z(11') = Q(11') -\int d3 d4 d5 t(134)g(45) {{\delta Z(51')} \over
{\delta K(3)}},
\end{equation}
Eqs(10)-(12) assume the form
\begin{equation}
\int d2 ( G_0^{-1}(12) -\Sigma(12))g(21') = \delta(1-1'),
\end{equation}
\begin{equation}
\Sigma(12) = -\int d3 t(132)L(3) +\int d3 d4 d5 t(134)g(45)\gamma(52,3),
\end{equation}
\begin{equation}
\gamma(12,3) = {\delta g^{-1}}(12)/ {\delta K(3)}.
\end{equation}
From Eqs.(17) and (19) follows, moreover, the equation for the vertex
\begin{equation}
\gamma(11',3) = \alpha(11';3) +\int d4 d5 \Theta(11',45) \gamma(45,3),
\end{equation}
with
\begin{equation}
\alpha(11',3) = {{\delta G_0^{-1}(11')} \over {\delta K(3)}} -\int d4 
{{\delta \Sigma(11')} \over {\delta L(4)}} \cdot {{\delta L(4)} \over {\delta
    K(3)}},
\end{equation}
\begin{equation}
\Theta(11',45) = \int d6 d7 {{\delta \Sigma(11')}\over {\delta g(67)}}
g(64)g(57).
\end{equation}
In Eq.(22) we have replaced the functional derivative of $g$ with respect
to $K$ by the vertex $\gamma$ using Dyson's equation which also leads
to a sign change.
The above equations are exact and hold both in the normal and the
superconducting state. In the normal state the two internal index pairs in 
$G(12)$ must obey either $p_1=q_2=0$ or
$p_2=q_1=0$. From Dyson's equation follows then
that the same holds for $Q(12)$ whereas the possible indices in $G_0(12)$,
$\Sigma(12)$, and $g(12)$ obey either $p_1=p_2=0$ or $q_1=q_2=0$.
In the superconducting state there are no such restrictions; the only
general requirement is that all these indices are associated with
fermionic operators. \par
The self-energy is a functional of $g$ and $L$, i.e., 
\begin{equation}
\Sigma = \Sigma[g,L].
\end{equation}
In order to derive the linearized equation for the anomalous
self-energy we split $\Sigma$ into the normal part $\Sigma_N$ and
a small anomalous part $\Sigma_{an}$. Expanding Eq.(23) up to linear
terms we obtain
\begin{equation}
\Sigma_{an}(11') = \int d2 d3 \Bigl({{\delta \Sigma(11')} \over {\delta
g(23)}}\Bigr)_N
\delta g(23) + \int d2 \Bigl({{\delta \Sigma(11')} \over {\delta 
L(2)}}\Bigr)_N \delta L(2),
\end{equation}
where $\delta g$ and $\delta L$ are of first order in the anomalous part
and the subscript $N$ means that the functional derivatives are to be
taken in the normal state. Now $\delta L$ has no linear contribution 
due to gauge invariance and thus 
drops out in Eq.(24). Calculating $\delta g$ from Dyson's equation Eq.(17)
in linear approximation we obtain
\begin{equation}
\Sigma_{an}(11') = \int d2 d3 \Theta(11',23) \Sigma_{an}(23),
\end{equation}
where $\Theta$ is given by Eq.(22) with all quantities taken in
the normal state. Eq.(25) represents the linearized equation for 
a general order parameter for superconductivity. In order to have
superconductivity the homogenous equation Eq.(25) must have a nonvanishing
solution for $\Sigma_{an}$. The highest temperature where this occurs is
the transition temperature $T_c$. \par
The basic quantitity to be calculated is according to Eq.(25) the
kernel $\Theta$. In order to find a functional equation for $\Theta$
we get from Eq.(18)
\begin{equation}
{{\delta \Sigma(11')} \over {\delta g(78)}} = \int d2 t(127) \gamma(81',2)
+\int d2 d3 d6 d7 d8 t(123)g(36){{\delta \gamma(61',2)} \over {\delta
g(78)}},
\end{equation}
and from Eq.(20)
$${{\delta \gamma(61',2)}\over {\delta g(78)}} = {{\delta \alpha(61',2)}
\over {\delta g(78)}} + $$
\begin{equation}
\int d9 d10 d9' d10'  \Theta(61',9'10')
{{\delta \gamma(9'10',2)}
\over  {\delta g(78)}} +\int d9' d10' {{\delta \Theta(61',9'10')} \over
{\delta g(78)}} \gamma(9'10',2).
\end{equation}
Solving Eq.(27) for $\delta \gamma/\delta g$, inserting the result into
Eq.(26) and inserting Eq.(26) into Eq.(22) yields an exact
functional equation for $\Theta$.

\section{ 1/N expansion for the kernel $\Theta$}

The equation for $\Theta$ obtained in the previous section is too difficult 
to be solved directly.
On the other hand it is exact and holds for any $N$ so it may serve as
a starting point for approximate treatments. In the following we
assume that $1/N$ may be used as a small parameter  and calculate
$\Theta$ up to order $1/N$. This will allow us to obtain the gap
equation Eq.(25) in leading order of the $1/N$ expansion. \par
The $N$-dependence of an equilibrium quantity is determined by
the number of coupling constants it contains and the number of
free internal summations. In equilibrium, i.e., without the 
source term $K$, we have, for instance,
\begin{equation}
g({{0q_1}\atop{\bar 1}}{{0q_2}\atop {\bar 2}})= \delta_{q_1q_2}
g({\bar 1}-{\bar 2}),
\end{equation}
\begin{equation}
\Sigma({{0q_1}\atop{\bar 1}}{{0q_2}\atop {\bar 2}})= \delta_{q_1q_2}
\Sigma({\bar 1}-{\bar 2}),
\end{equation}
\begin{equation}
\gamma({{oq_1}\atop {\bar 1}}{{oq_2} \atop{\bar 2}},{{p_3q_3} \atop {\bar 3}})
={-1 \over N}\delta_{p_3q_3}\delta_{q_1q_2} \gamma_c({\bar 1}{\bar 2},{\bar
  3}) + \delta_{q_1p_3} \delta_{q_2q_3} \gamma_s({\bar 1}{\bar 2},{\bar 3}),
\end{equation}
and similar expressions for $\alpha$ and $\Theta$. 
In Eqs.(28)-(30) all the indices $q_1,q_2,...$ are assumed to be larger than
zero. The equilibrium Green's function $g$, the self-energy $\Sigma$,
the charge vertex $\gamma_c$ and the spin vertex $\gamma_s$ are free of
internal indices. In the following we need only the leading O(1)
contributions for these quantities. From Eqs.(13), (17) and (19) follows
that the spin vertex becomes then simply equal to $\delta({\bar 1}-{\bar 2})
\delta({\bar 1}-{\bar 3})$. Using the constraint Eq.(4) one also
recognizes that the charge vertex $\gamma_c$ defined in Eq.(30) is
equal to the element $q_1=q_2>0,p_3=q_3=0$ of the general vertex $\gamma$
on the left-hand side of Eq.(30) which is the motivation to use
prefactors in Eq.(30) in defining $\gamma_c$. \par

Decomposing the labels $1$,$1'$, etc. into their internal and external
parts the anomalous self-energy or order parameter has the form
$\Sigma_{an}({{p_10} \atop {\bar 1}} {{0q_1'} \atop {\bar 1}'})$, or in
more detail, $\Sigma_{an}({{\sigma_1 m_1 0} \atop {\bar 1}} 
{{0 \sigma_1' m_1'} \atop {\bar 1}'})$. In the following we are interested
in order parameters which are structureless in the flavor indices, i.e.,
we assume always $m_1=m_1'$. With respect to spin indices we consider
either singlet or triplet pairing. Both cases are included if we put
$\sigma_1=-\sigma_1'$. 
Correspondingly we will write for the internal indices $p_1,{\bar p}_1$
thus indicating the special relationship between these two indices.
Exposing explicitly the internal indices the gap equation Eq.(25) becomes 
\begin{equation}
\Sigma_{an}({{p_10} \atop {\bar 1}}{{0{\bar p}_1} \atop {\bar 1}'}) =
\sum_{p_2} \int d2 d3 \Theta({{p_10} \atop {\bar 1}} {0{\bar p_1}
\atop {\bar 1}'}, {{p_20} \atop {\bar 2}} {{0{\bar p}_2} \atop {\bar 3}})
\Sigma_{an} ({{p_20} \atop {\bar 2}} {{0{\bar p}_2} \atop {\bar 3}}),
\end{equation}
where $\Theta$ is to be taken in the normal state. Eq.(31) also 
makes visible which indices for $\Theta$ are actually needed
in the gap equation. 
A closer examination shows that the first two contributions on the
right-hand side of Eq.(27) are of higher orders in $1/N$ than the third
term in this equation for the needed combination of indices.. 
Dropping these terms and inserting Eqs.(26) and (27)
into Eq.(22) yields the following equation for $\Theta$ valid up to
O(1/N) and for the above combinations of internal indices:

$$\Theta(11',910) = \int d2 d7 d8 t(127) \gamma(81',2)g(79)g(108)  +$$
\begin{equation}
\int d2 d3 d6 d7 d8 d9' d10' t(123) g(36) {{\delta \Theta(61',9'10')}
\over {\delta g(78)}} \gamma(9'10',2) g(79) g(108).
\end{equation}
The first term on the right-hand side of Eq.(32) could have been also
obtained directly
from Eq.(18). Considering only anomalous contributions to $\Sigma$
the first term on the right-hand side of Eq.(18) cannot contribute,
wheras the second one can contribute in two ways: a) via an anomalous $g$
and a normal vertex $\gamma$, b) via a normal $g$ 
and an anomalous vertex $\gamma$. Case a) corresponds to
the usual situation of Eliashberg theory and also of the slave boson
treatment of superconductivity in the $t-J$ model in O(1/N): $\Sigma_{an}$
consists then of a Fock diagram containing an anomalous Green's function 
and an effective interaction taken in the normal state. Linearizing $g$ in 
$\Sigma_{an}$
case a) immediately yields the first term on the right-hand
side of Eq.(32). Its evaluation is straightforward and shows that it is of
O(1/N) for the relevant combination of internal indices. This indicates
that $T_c \rightarrow 0$ for
$N \rightarrow \infty$ which agrees with the fact that at $N=\infty$ the 
system consists of renormalized, but non-interacting fermions. \par 
The leading contribution of the second term on the right-hand side of 
Eq.(32) is obtained by taking the spin-flip part in the hopping matrix 
element $t$ and the spin vertex $\gamma_s$ in O(1) for the vertex $\gamma$.
As a result only one sum over internal
indices survives.The second term in Eq.(32) is of
$O(1/N)$ if the two functions
\begin{equation}
g^{(1)}({\bar 6}{\bar 1}',{\bar 2}|{\bar 7}{\bar 8}) =
\sum_{p_3} {{\delta \Theta\Bigl({{p_30} \atop {\bar 6}}{{0{\bar p}_1} \atop
{\bar 1}'}, {{0p_1} \atop {\bar 2}} {{0p_3} \atop {\bar 2}}\Bigr)} 
\over {\delta g\Bigl(
{{{\bar p}_10} \atop {\bar 7}}{{0p_1} \atop {\bar 8}}\Bigr)}},
\end{equation}
\begin{equation}
g^{(2)}({\bar 6}{\bar 1}',{\bar 2}|{\bar 7}{\bar 8}) =
\sum_{p_3} {{\delta \Theta\Bigl({{p_30} \atop {\bar 6}}{{0{\bar p}_1} \atop
{\bar 1}'}, {{p_30} \atop {\bar 2}}{{p_10} \atop {\bar 2}} \Bigr)} 
\over {\delta g\Bigl(
{{{\bar p}_10} \atop {\bar 7}}{{0p_1} \atop {\bar 8}}\Bigr)}},
\end{equation}
are of O(1). Equations for $g^{(i)},i=1,2$ can be obtained from Eq.(32) by
taking appropriate derivatives. An examination of the
various terms shows that
terms with a second functional derivative of $\Theta$ are smaller by a
factor $1/N$ compared to the leading ones and thus can be neglected.
One then finds that the functions $g^{(i)}$ have the form
\begin{equation}
g^{(1)}({\bar 6}{\bar 1}',{\bar 2}|{\bar 7}{\bar 8}) = \delta({\bar 2}-
{\bar 8}) g^{(1)}({\bar 6}{\bar 1}',{\bar 2}|{\bar 7}),
\end{equation}
\begin{equation}
g^{(2)}({\bar 6}{\bar 1}',{\bar 2}|{\bar 7}{\bar 8}) = \delta({\bar 2}-
{\bar 7}) g^{(1)}({\bar 6}{\bar 1}',{\bar 2}|{\bar 8}).
\end{equation}
The reduced functions $g^{(i)}({\bar 6}{\bar 1}',{\bar 2}|{\bar 8})$
(the use of the same name for corresponding functions with different
number of arguments should not cause any confusion) satisfy the integral
equation
\begin{equation}
g^{(i)}({\bar 1},{\bar 1}',{\bar 2}|{\bar 7}) = 
h^{(i)}({\bar 1},{\bar 1}',{\bar 2}|{\bar 7}) -
\int d{\bar 3} d{\bar 4} d{\bar 6} Nt({\bar 1}{\bar 3}{\bar 4})
g({\bar 6}{\bar 4})g^{(i)}({\bar 6}{\bar 1}',{\bar 3}|{\bar 7}) 
g({\bar 2}{\bar 3}),
\end{equation}
with
\begin{equation}
h^{(1)}({\bar 1}{\bar 1}',{\bar 2}|{\bar 7}) = Nt({\bar 1}{\bar 1}'
{\bar 7}) g({\bar 2}{\bar 1}'),
\end{equation}
\begin{equation}
h^{(2)}({\bar 1}{\bar 1}',{\bar 2}|{\bar 7}) = -\int d{\bar 3} d{\bar 4}
Nt({\bar 1}{\bar 3}'
{\bar 4}) \gamma_c({\bar 7}{\bar 1}',{\bar 4}) g({\bar 2}{\bar 3}).
\end{equation}
Here $t({\bar 1}{\bar 2}{\bar 3})$ denotes the external part of
$t(123)$,i.e., the first three factors outside of the bracket in Eq.(9). 
Inserting the explicit expression for $t$ into Eq.(37) one finds that
Eq.(37) represents an integral equation for $g^{(i)}$ with a kernel
consisting of not more than 6 separable contributions. The solution of Eq.(37) 
thus reduces to the inversion of at most 6x6 matrices similar as in the
case of the charge vertex $\gamma_c$ dicussed in Ref. \cite{Zeyher4}
Inserting the
resulting expressions back into Eq.(32) and performing Fourier transforms
one obtains after some elementary, but tedious algebra the following results.
The linearized gap equation Eq.(25) becomes 
\begin{equation}
{\Sigma}_{an}(k) = -{T \over {NN_c}}\sum_{k'}\Theta(k,k')
{1 \over {\omega_{n'}^2 +\epsilon^2({\bf k'})}}
{\Sigma}_{an}(k'). 
\end{equation}
$N_c$ is the number of cells and k the supervector $k=(n,{\bf k})$,
where $n$ denotes a fermionic Matsubara frequency $\omega_n=(2n+1)\pi T$
with $T$ being the temperature. $\epsilon({\bf k})$ is
the one-particle energy with momentum $\bf k$ in the limit
$N \rightarrow \infty$ given by the implicit equation
\begin{equation}
\epsilon({\bf k}) = {{\delta}\over 2} t({\bf k}) -J({\bf k})\cdot
{1 \over N_c} \sum_{\bf p}cos(p_x) f(\epsilon({\bf p}-\mu))
\end{equation}
where $\delta$ is the doping, $t({\bf k})$ and $J({\bf k})$ the
Fourier transforms of the coupling constants $t_{ij}$ and $J_{ij}$,
respectively, $\mu$ is the chemical potential and $f$ the Fermi
function. \par
The kernel $\Theta$ in Eq.(40) 
consists of four different
terms 
\begin{equation}
\Theta(k,k') = \Theta^{(1)}(k,k')+\Theta^{(2)}(k,k')+\Theta^{(3)}(k,k')
+\Theta^{(4)}(k,k').
\end{equation}
The first two terms are given by 
\begin{equation}
\Theta^{(1)}(k,k') = \mp t({\bf k}') \mp J({\bf k}-{\bf k}')
-t({\bf k}') -J({\bf k}-{\bf k}'),
\end{equation}
\begin{equation}
\Theta^{(2)}(k,k') = (t({\bf k}')+J({\bf k}-{\bf k}'))(\gamma_c(k,k'-k)+1).
\end{equation}
Here, $\gamma_c$ is the charge vertex in O(1) determined by
\begin{equation}
\gamma_c(k,q)=-1 + \sum_{\alpha,\beta =1}^6 F_\alpha(k)
(1+\chi(q))^{-1}_{\alpha \beta} \chi_{\beta 2}(q),
\end{equation}
with the susceptibility matrix
\begin{equation}
\chi_{\alpha \beta}(q) = {1 \over N_c} \sum_{\bf k} E_\alpha({\bf k},{\bf q})
\; F_\beta({\bf k})\;\;{ {f(\epsilon({\bf k}+{\bf q}))-f(\epsilon({\bf k}))}
\over {\epsilon({\bf k}+{\bf q}) -\epsilon({\bf k}) -i \nu_n}}.
\end{equation}
$q$ is the supervector $q=({\bf q}, i\nu_n)$ where $\nu_n$ denotes
the bosonic Matsubara frequency $2\pi n T$. The two vectors E and F are
given by
\begin{equation}
E_\alpha({\bf k},{\bf q}) = (1,t({\bf k}+{\bf q})+J({\bf q}),cosk_x,sin k_x,
cosk_y, sink_y),
\end{equation}
\begin{equation}
F_\beta({\bf k}) = ( t({\bf k}),1,Jcosk_x,Jsink_x,Jcosk_y,Jsink_y).
\end{equation}
The explicit expressions for the third and fourth contributions to 
$\Theta$ are 
\begin{equation}
\Theta^{(3)}(k,k') = -\sum_{r=1}^5 \tilde{E}_r(-{\bf k}) 
\tilde{\chi}_{2r}(k-k') \gamma(k,k'-k) + \sum_{r,s=1}^5 \tilde{E}_s
(-{\bf k}) \tilde{\chi}_{rs}(k-k') \tilde{V}_r(k-k'),
\end{equation}
\begin{equation}
\Theta^{(4)}(k,k') = \pm \sum_{r,s=1}^5 (
\tilde{E}_s(-{\bf k}) \tilde{\chi}_{rs}(k+k'))(\tilde{F}_r(-{\bf k}')          + \tilde{W}_r(k+k')),
\end{equation}
with the vectors $\tilde{V}$ and $\tilde{W}$ 
\begin{equation}
\tilde{V}_r(k-k') = \sum_{s=1}^5 (1+\tilde{\chi}(k-k'))^{-1}_{rs}
\tilde{\chi}_{2,s}(k-k') \gamma_c(k,k'-k),
\end{equation}
\begin{equation}
\tilde{W}_r(k+k') = -\sum_{s,t=1}^5 (1+\tilde{\chi}(k+k'))^{-1}_{rs}
\tilde{F}_t(-{\bf k}')\tilde{\chi}_{ts}(k+k').
\end{equation}
The vectors $\tilde{E}$ and $\tilde{F}$ have five components and
are obtained from the six-component vectors $E$ and $F$ by omitting
the second component. In an analogues way, $\tilde{\chi}_{2,s}$ is 
obtained from the second row of the matrix $\chi$ by omitting the 
second element. Similarly, the 5x5 matrix $\tilde{\chi}$ is obtained from
the 6x6 matrix $\chi$ by dropping the second row and second column. 
The upper (lower) signs in the contibutions $\Theta^{(i)}$ refer
to singlet (triplet) pairings.\par
The kernel $\Theta(k,k')$ is invariant under the 
transformations of the point group $C_{4v}$ of the square lattice.
As a result $\Sigma_{an}(k)$ transforms corresponding to one of the
five irreducible representations $\Gamma_i$ of $C_{4v}$, and it can be
chosen to be either even or odd in Matsubara frequencies.
$\Gamma_1$ corresponds to singlet
s- or extended s-wave, $\Gamma_3$ to singlet d-wave- and $\Gamma_5$
to triplet p-wave pairing for even frequency pairing. 
The linearized gap equation Eq.(40) splits 
completely into its irreducible parts. This also means that the
momenta ${\bf k},{\bf k'}$ can be restricted to 1/8 of the Brillouin zone
which greatly simplifies the numerical solution of Eq.(40).
In the following we will deal with even frequency pairing unless
the opposite is explicitly stated and drop for simplicity the index which 
differentiates
between even and odd frequency-pairing. The irreducible kernels and
order parameters will thus be denoted by $\Theta_i(k,k')$
and $\Sigma_{an,i}(k)$, respectivley. \par

\section{Discussion of the gap equation and numerical results}

According to Eq.(42) the total kernel $\Theta$ is the sum of four
contributions. The first one, $\Theta^{(1)}$, represents the instantaneous
part of $\Theta$. From Eq.(40) follows that positive values for $\Theta^{(1)}$
correspond to repulsion, negative values to attraction between electrons. 
Keeping only $\Theta^{(1)}$ and decomposing it into its irreducible
symmetry components the gap equation Eq.(40) becomes for the representation
i=1...5 of the point group $C_{4v}$ of a square lattice
\begin{equation}
\Sigma_{i,an}({\bf k}) = {{-1}\over{NN_c}} \sum_{\bf k'} \Theta_i^{(1)}
({\bf k},{\bf k}') \chi({\bf k}') \Sigma_{i,an}({\bf k}'),
\end{equation}
with 
\begin{equation}
\chi({\bf k}') = {tanh((\epsilon({\bf k}')-\mu)/T)\over {2(\epsilon({\bf k}')
-\mu)}},
\end{equation}
$\Theta^{(1)}_2=\Theta^{(1)}_4=\Theta^{(1)}_5=0$, and
\begin{equation}
\Theta^{(1)}_1({\bf k},{\bf k}') = 8|t|\eta_1({\bf k}') -8J\eta_1({\bf k}) 
\eta_1({\bf k}'),
\end{equation}
\begin{equation}
\Theta^{(1)}_3({\bf k},{\bf k}') = -8J\eta_3({\bf k}) \eta_3({\bf k}').
\end{equation}
In Eqs.(55)and (56) we assumed nearest-neighbor hopping with $t_{ij}=-|t|$
and used the basis functions $\eta_1({\bf k})=(cos(k_x)+cos(k_y))/2,
\eta_3({\bf k})=(cos(k_x)-cos(k_y))/2$. Inserting the kernels Eqs.(55) and (56)
into the gap equation yields the following conditions for a finite $T_c$:
\begin{equation}
s-wave \;\; pairing \;(\Sigma_{an} \sim constant): 1+a|t| +{{ab_1|t|J} \over
{1-Jb_1}}=0,
\end{equation}
\begin{equation}
extended \;\;s-wave \;\;pairing \;(\Sigma_{an} \sim \eta_1({\bf k})):
1-b_1 + {{ab_1|t|J} \over {1+a|t|}} = 0,
\end{equation}
\begin{equation}
d-wave \;\;pairing \;(\Sigma_{an} \sim \eta_3({\bf k})):
1-b_3 J=0.
\end{equation}
Here we used the following abbreviations
\begin{equation}
a = {8 \over{NN_c}} \sum_{\bf k'}\chi({\bf k'}) \eta_1({\bf k}'),
\end{equation}
\begin{equation}
b_l = {8 \over{NN_c}} \sum_{\bf k'}\chi({\bf k}') \eta_l^2({\bf k}'),
\end{equation}
for $l=1,3$. A numerical evaluation shows that in the interesting
parameter region $0 \leq J \leq 0.3, 0 \leq \delta \leq 0.8$
($\delta$ is the doping) and at low temperaturs
$a,b_1,b_3$ are positive and $b_1 < 3$. This means that there is never
an instability with respect to constant s-wave pairing. $b_1$ is an 
increasing, $b_3$ a strongly decreasing function with increasing $\delta$
and these two functions cross around $\delta \sim 0.6$. Thus
d-wave pairing is stable in any case for $\delta < 0.5$. For
$0.5 \leq \delta \leq 0.8$ $a|t|$ is much larger than 1 so that
the third term in Eq.(58) cancels the second term  to a large extent
making extended s-wave pairing also in this region unfavourable. Thus
we find that the instantaneous contribution to the kernel strongly
favours d-wave pairing in the interesting parameter regime. The
reason for this is to a large extent a band structure effect: $b_3$
is much larger than $b_1$ at small dopings because of large contributions
around the X-point. At larger doping the competing extended s-wave
pairing is strongly suppressed by its coupling to the hopping term,
which ultimately is caused by the constraint. \par    
$\Theta^{(2)}$, $\Theta^{(3)}$, and  $\Theta^{(4)}$ are retarded 
contributions to $\Theta$. 
$\Theta^{(2)}$ is mainly determined by collective charge fluctuations 
due to the poles of $\gamma_c$. $\Theta^{(3)}$ and $\Theta^{(4)}$ originate
from the anomalous part of the vertex and involve both
charge and spin fluctuations. The latter dominate in $\Theta^{(4)}$
in agreement with the sign change between its singlet and triplet
contribution. Quantitatively, $\Theta^{(4)}$ is by far the largest 
of the retarded contributions. 
Our $\Theta$ does not contain terms which are related to the magnons
of the undoped case. Such contributions are of 
higher order in $1/N$ than those considered above and are thus neglected.
\par
Our expression for $\Theta$ is different from that of the slave boson 
approach \cite{Grilli1}. 
$\Sigma_{an}$ in the slave boson approach essentially consists
of a spinon Fock term with an effective interaction taken in the normal
state times an anomalous spinon Green's function. Such an contribution
clearly corresponds to the case a) discussed after Eq.(32), i.e.,
to the first term on the right-hand side of Eq.(32), or, equivalently.
to the sum of $\Theta^{(1)}$ and $\Theta^{(2)}$. Our contributions
$\Theta^{(3)}$ and $\Theta^{(4)}$ have no analogue in the 
slave boson approach though they are clearly also of O(1/N). The
differences between the two approaches can be made more explicitly by
considering two limiting cases. For $J \rightarrow 0$ $\Theta$ reduces
to Eq.(9) of Ref. \cite{Greco1}
 The underlying interactions are still spin-dependent
as can be seen from the singlet versus triplet case. Such a 
dependence on spin is not present in the corresponding slave boson 
expression \cite{Kotliar1}. 
Furthermore, the high-frequency limit of $\Theta$, i.e.,
$\Theta^{(1)}$ should be identical to that of the slave boson expression
if the two approaches are equivalent. 
The slave boson result for $\Theta^{(1)}$
differs, however, from Eq.(43): the argument in one of the two hopping terms 
is replaced by $\bf k$ and there is an additional term
proportional to $1/\delta$. Eq.(5) agrees, however, with previous
X-operator results based on the diagrammatic methods for X-operators,
for instance, Eq.(35) of Ref.\cite{Sandalov1}. The expression for
$\Theta^{(1)}$ of Ref. \cite{Onufrieva1} contains only one of the two
hopping terms $t({\bf k}')$ of Eq.(5) which, we think, is incorrect.
The general question then arises why is it possible to obtain different
O(1/N) expressions for the same quantitiy in the two approaches using
the same Hamiltonian. The differences arise because 
different Hilbert spaces are used in the two cases. This difference is
already indicated in the definition of the order parameters for 
superconductivity. In the X-operator approach all vectors of the
Hilbert space are eigenvectors of the constraints $Q_i$ with eigenvalue
$N/2$. Expectation values of operators are only nonzero if, in the
slave boson language, 
the number of created slave particles is equal to
the number of annihilated slave particles in these operators. The 
superconducting order parameter of slave boson theory, on the other hand,
 consists of the 
expectation values $<b>$ and $<c^\dagger c^\dagger>$. Clearly, 
these expectation values have no analogues in the X-operator approach.

\subsection{Eigenvalues of $\Theta$ in the static limit}

In the weak-coupling case it is often assumed that the solution of the
gap equation Eq.(40) is of BCS-type for each symmetry $\Gamma_i$, i.e.,
\begin{equation}
T_{ci} = 1.13 \omega_c e^{1/ \lambda_i}.
\end{equation}
$\omega_c$ is a suitable cut-off and $\lambda_i$
the smallest eigenvalue of a matrix which is essentially equal to
the static limit of the kernel
$\Theta$ and given by Eq.(10) of Ref. \cite{Greco1}. This matrix includes also
the prefactor $1/N$ on the right-hand side of Eq.(40) and we put from now on
$N$ always equal to $2$. Actually we will find that Eq.(62) is a rather
poor approximation for $T_c$ because $\Theta$ contains instantaneous 
and retarded
contributions which have different cut-offs. Nevertheless the study of the
static kernel $\Theta$ and its lowest eigenvalues has traditionally played
a great role in discussing superconductivity in t-J models. \par
Fig. 1a) shows the terms $\Theta^{(1)}$ and $\Theta^{(2)}$ and Fig. 1b)
the terms $\Theta^{(3)}$,$\Theta^{(4)}$ and $\Theta$ for singlet pairing
at zero frequencies for a fixed first momentum 
$\bf k$ = (2.465,0.309) as a function of the second
argument $\bf k'$. $\bf k '$ moves counterclockwise around the Fermi line
passing through the points $X,Y ({\bar X},{\bar Y})$ along the positive
(negative) x- and y-axis, respectively. Note that the letters $X,Y$
denote not the ${\bf k}$-points $(\pi,0),(0,\pi)$ but 
the points on the Fermi line between the $\Gamma$-point and 
the points $(\pi,0),(0,\pi)$, respectively.
The doping is $\delta = 0.17$
and $J=0.3$. We used 22 $\bf k$-points along 1/8 of the Fermi line
and a net of 300x300 $\bf k$-points in the Brillouin zone. 
Positive values of $\Theta$ mean repulsion, negative ones
attraction between electrons in the s-wave channel.
 $\Theta^{(1)}$ and $\Theta^{(2)}$ 
originate from the normal, $\Theta^{(3)}$ and $\Theta^{(4)}$ from
the anomalous vertex. $\Theta^{(1)}$ is due to instantaneous, the
other terms due to retarded interactions. The Figures show that
$\Theta^{(4)}$ and $\Theta^{(1)}$ are by far the largest contributions.
Both are dominated by the d-wave component and add up in a coherent way.
$\Theta^{(4)}$ has different signs for singlet and triplet pairings.
This as well as the explicit calculation shows that important contributions
to $\Theta^{(4)}$ come from spin fluctuations within the partially filled
band. There are no spin contributions related to the Heisenberg
term and the magnon spectrum at zero doping because these terms are at
least one order in 1/N smaller than the considered ones. $\Theta^{(3)}$
is due to charge (and spin) excitations and it is very small. Finally,
$\Theta^{(2)}$ is due to charge fluctuations and consists of  an attractive
s-wave component (which, however, is cancelled by a repulsive
s-wave term in $\Theta^{(1)}$) and a d-wave part which cancels partially
that of the $\Theta^{(1)}$ term. Figs.1 demonstrate the importance
of anomalous vertex contributions to $\Theta$,
the d-wave character of the leading contributions, and the competition
between instantaneous and retarded interactions. \par
Fig. 2 shows the lowest eigenvalues $\lambda_i$ for each of the
five representations $\Gamma_i$. In the calculation we used 5 $\bf k$-points
along 1/8 of the Fermi line and a net of 300x300 $\bf k$-points in the
Brillouin zone. The eigenvalues $\lambda_i$ decrease 
with decreasing 
doping $\delta$ and diverge at $\delta \sim \delta_{BO} \sim 0.13$.
At this doping value one of the six eigenvalues of the 6x6 matrix $1+\chi$ 
in Eq.(45) goes through zero . As a result there is a soft mode 
which freezes into a static incommensurate bond-order wave of d-wave symmetry
for $\delta < \delta_{BO}$. This instability causes a divergence in
some of the contributions to $\Theta$ at $\delta_{BO}$ and, as a precursor,
the large negative values seen in Fig. 2 for $\delta  \leq 0.15$. The solid
line in Fig. 2 exhibits the lowest eigenvalue $\lambda_3$ associated 
with d-wave pairing. Its absolute magnitude is much larger than all
the other eigenvalues and this is true for all dopings. Inserting
$\lambda_i$ into Eq.(40) it is clear that superconducting 
instabilities for non-d-wave symmetries are more of academic interest
because the corresponding $T_c's$ would be extremely small. One
concludes that the normal state is generally unstable against
superconductivity in every symmetry channel but that only the d-wave
$(\Gamma_3)$ instability is strong enough to account for high
transition temperatures. \par  

\subsection{Frequency dependence of $\Theta$}

For either even or odd frequency pairing the frequency dependence of
$\Theta$ can be rewritten in terms of one frequency argument $\omega_n$ which
is equal to the 
difference of the two original
frequency variables. In order to illustrate the frequency dependence
of the retarded part of $\Theta$, $\Theta^{ret}$, in the case of d-wave
scattering, we average the momenta
in $\Theta^{ret}$ over the Fermi line in the irreducible Brillouin
zone using the eigenvector of the lowest eigenvalue $\lambda_3$. Finally we
perform the analytic continuation $i \omega_n \rightarrow \omega + i
\eta$. \par
Fig.3 shows the negative imaginary part of $\Theta^{ret}_3(\omega +i \eta)$
for $\eta = 0.005$. It
is the analogue of the familiar function $\alpha^2F(\omega)$ of
Eliashberg theory for phonon-induced s-wave superconductivity. In our case
this function is no longer positive-definite. This is a result of
the projection of the total kernel on the $\Gamma_3$ symmetry.   
This projection involves a sum of $\bf k'$ over the 8 symmetry-related 
points along the Fermi line with factors determined by the representation 
$i$. In the case of d-wave pairing these factors are positive for small
and large momentum transfers (of the order or a reciprocal lattice vector)
and negative for intermediate momentum transfers of about half 
of a reciprocal lattice vector. For collective density fluctuations 
described by $\Theta^{(2)}$ 
this means that high-frequency contributions appear with a negative and
low-frequency contributions with a positive sign in the d-wave kernel.
A similar behavior is found for the other contributions. As a result
$-Im \Theta_3^{ret}$ is positive for $\omega \leq \omega_2 \sim |t|$
and negative for $\omega \geq \omega_2$. Using a momentum average 
with a constant weigth of $\Theta^{ret}$ corresponding to constant
s-wave pairing  would yield a curve which is similar to that in Fig.3 for
$\omega \leq \omega_2$ but opposite in sign for $\omega \geq \omega_2$.
$-Im \Theta_3^{ret}$ describes the spectral distribution and spectral
weight of spin and charge excitations involved in d-wave scattering.
This function extends over a wide frequency region of about $2|t|$.
Its high-frequency part is typical for collective charge fluctuations.
It also contains substantial spectral weight 
at lower frequencies exhibiting a rather linear increase in frequency at 
the low-frequency end of the spectrum. 
Fig. 4 shows the corresponding real part of $\Theta_3^{ret}$. It is 
weakly negative at low frequencies up to about $\omega_1 \sim J/2$,
changes then from negative to positive values until about $\omega \sim
\omega_2$. The retarded interaction between electrons is thus attractive for
$0 \leq \omega \leq \omega_1$ and strongly repulsive for $\omega_1 
\leq \omega \leq \omega_2$. Somewhat above $\omega_2$ the real part 
of $\Theta_3^{ret}$ jumps to large negative values and approaches zero
from below in the high-frequency limit. The inset of Fig. 4 shows the real
part of $\Theta_3$ at small frequencies with a larger resolution using
$\eta = 0.002$.

\subsection{Transition temperatures}

Figs. 1a) and b) show that the instantaneous and the retarded contributions
to $\Theta$ are of similar magnitude and often compete with each other. 
In the calculation of $T_c$ they are associated with quite different
cut-offs. The instantaneous part is characterized by a cut-off determined
by the width of the effective band whereas the cut-off relevant for
the retarded part is set by the frequency range where attraction dominates
characterized roughly by $J$. In view of these complications we 
developed a method to solve Eq.(40) directly, avoiding the use
of pseudopotentials. The only simplification
we use is to put the momenta of the retarded kernel on the Fermi line.
The validity of this approximation has been checked numerically and 
holds very well in
our case. Using the fact that the instantaneous kernel consists only
of a few separable contributions Eq.(40) can be reduced to a linear matrix
problem. The number of rows and columns are given by the number of 
considered Matsubara frequencies times the number of $\bf k$-points on the 
Fermi line in the irreducible Brillouin zone. Numerical tests showed that
well converged results can be obtained with about 300 Matsubara frequencies
and 5 $\bf k$-points along 1/8 of the Fermi line down to 
values for $T_c$ of $\sim 0.002$. Details of our method will be given 
elsewhere. \par

The squares in Fig. 5 (joined by a solid line) show $T_c$ for 
$\Gamma_3$-symmetry for doping $\delta > \delta_{BO}$. The broken line
exhibits $T_c$ if only the instantaneous part of the kernel is used. The
dotted line corresponds to $T_c$ if the charge-charge contribution
(last term in Eq.(1)) is dropped. Finally the dashed-dotted line 
describes $T_c$ if only the retarded kernel is taken into account.
\par
In order to understand the curves in Fig. 5 we consider the real part
of the retarded kernel as a function of frequency, as shown
in Fig. 4. Taking only the part between zero and $\omega_2$
into account would yield
rather large values for $T_c$. To realize this one can decompose the
real part into a large, constant repulsive part between
$\omega=0$ and $\omega = \omega_2$ plus the difference which is
non-zero and attractive at low frequencies. Changing the cut-off from
$\omega_2$ to $\omega_1$ using a pseudopotential description decreases
strongly the effective repulsion yielding a large net attraction
between 0 and $\omega_1$ and thus a high value for $T_c$. Model calculation 
show that the
large negative part in the real part of $\Theta_3$ above $\omega_2$
is very harmful to superconductivity: Lowering first the cut-off from
$\sim 3|t|$ to $\omega_2$ increases the effective potential, reducing 
further the cut-off to $\omega_1$ decreases again the effective potential.
The overall result is a rather modest attraction
between $\omega=0$ and $\omega=\omega_1$. This explains the rather
low values for $T_c$ calculated from the retarded part of $\Theta$
alone, as shown by the dash-dotted line in Fig. 5. An alternative
consideration would start from the instantaneous part in $\Theta$
leading to the transition temperatures shown by the broken line
in Fig. 5. Using a realistic value for the energy unit of about $8000 K$
(note that the usual effective hopping $t_{exp}$ corresponds according
to Eq.(1) to $|t|/2$)  these values are already of the order of $100 K$
and comparable to the experimental ones. Adding the retarded part
of the kernel does not lead to a further increase in $T_c$ but rather
to a slight decrease as illustrated by the squares in Fig. 5. 
Model calculations show that a purely attractive $\Theta^{ret}$
would always increase $T_c$. The observed lowering of $T_c$ must therefore be
due to the repulsive part in $\Theta^{ret}$ above $\omega_1$. This part
plays a role because the $T_c$ due to the instantaneous part alone
is large enough to couple strongly the frequencies above and below $\omega_1$
in the gap equation. The doping dependence of $T_c$ is determined by 
various processes
which compete with each other. For instance, the density of states
decreases with doping which decreases the effective coupling. On the 
other hand , the energy scale for $T_c$ is in the case of the instantaneous
term set by the effective band width which increases with doping. Fig. 5
shows that the net effect of these and other processes is to lower
$T_c$ substantially with increasing doping for 
$\delta >\delta_{BO}$. 
The eigenvalue $\lambda_3$ of the static kernel assumes according to Fig. 2
large negative
values near $\delta_{BO}$ anticipating the incipient bond-order
wave instability. The associated soft mode does not affect the instantaneous
but the retarded part. Fig. 5 shows that the dashed line indeed increases
steeply approaching $\delta_{BO}$ from above. On the other hand little
effects are seen in the squares representing the full calculation. 
The reason for this is that at the low $T_c$'s of the dashed curve
the relevant frequencies
lie in the attractive region of $\Theta^{ret}_3$ whereas for the high $T_c$
values of the squares these frequencies lie already in the repulsive 
part. From this one may conclude that neither soft modes nor large
negative eigenvalues of the static kernel guarantee large $T_c$ values.
In particular, there is no simple connection between $T_c$ and $\lambda_3$.
\par
The squares (connected by a dashed line) in Fig. 6 show 
$T_c$ as a function of the exchange constant $J$ for a doping $\delta = 0.17$.
$T_c$ approaches 0 for $J \rightarrow 0$ in agreement with previous
findings \cite{Greco1}. 
Except at small values of $J$ $T_c$ depends linearly on $J$.
This is quite in contrast to the BCS-formula with an exponential 
dependence on $J$. The main reason for this is that the
instantaneous term plays a major role and the momenta in it cannot
confined to the Fermi surface in calculating $T_c$. Eq.(53) is then
a more appropriate formula for $T_c$. It is, however, too simple to argue
that $b_3$ in that formula is rather independent on $J$ and $\sim 1/T$
for $T \geq 0.01$. The rather perfect linear behavior is the result
of more subtle dependencies such as the $J$-dependence of the one-particle
energies and the non-rigidity of the bands as a function of $J$. \par
It has been argued \cite{Feiner1,Riera1}
that the nearest-neighbor Coulomb potential $V$ is
not neglegible in the cuprates. Adding this term
\begin{equation}
H' = {V \over {2N}} \sum_{{<ij>}\atop {p,q=1...N}} X_i^{pp} X_j^{qq},
\end{equation}
to the Hamiltonian Eq.(1) we have calculated $T_c$ as a function of $V$.
The circles (connected by a solid line) in Fig. 6 present the result for
$J=0.3$ and $\delta = 0.17$. $T_c$ decreases strongly with increasing $V$
and is extremely small for $V \geq 3J/4$. This may be understood by looking
just at the instantaneous term: Eq.(63) yields an additional term 
$2V({\bf k}-{\bf k}')$ on the right-hand side of Eq.(43), canceling the 
$J$-terms exactly for $V=J$. $T_c$ would thus vanish if only the 
instantaneous term would be present. For $V=J/2$ $H'$ cancels 
the charge term in Eq.(1). According to Fig. 6 this means a drop of $T_c$
by about a factor 5 compared to the value at $V=0$ which also agrees
with Fig. 5. Such a big drop of $T_c$ due to the charge term of the $t-J$
model seems very surprising because the latter can only produce
effects $\sim \delta^2$ and is therefore often omitted. Our calculation
shows that this term cannot be neglected in a calculation of $T_c$
and increases $T_c$ substantially. \par  
Fig. 7 shows $T_c$ as a function of $\delta$ for $J=0.3$ and $V=0.15$.
Due to the Coulomb repulsion the $T_c$ values are rather low.
The important frequencies in the solution of the gap equation are
also low and mainly located in the attractive region of the retarded 
kernel, Adding the retarded to the instantaneous part thus increases
$T_c$. The two curves in Fig. 7 demonstrate this effect. \par
Figs. 8 and 9 show the influence of a second-nearest neighbor hopping 
term $t'$ on $T_c$. In these figures a lower value of 0.2 
has been chosen for $J$ similar as in Ref. \cite{Zeyher1} 
in order to have a reasonably
low $\delta_{BO}$ and a Van Hove singularity not too far away from
optimal doping. Assuming a linear dependence of $T_c$ on $J$ the
the absolute values for $T_c$ are similar in the corresponding Figs. 5 
and 8. $T_c$ is thus rather robust to changes in $t'$. $T_c$ decreases in 
Fig. 8 somewhat slower than in Fig. 5 due to the larger density of states
in the surroundings of the Van Hove singularity. The difference between
squares and circles is also larger in Fig. 8 and decreases much slower
with increasing doping. These effects are caused by the retarded part
of the kernel which in Fig. 8 is less attractive below $\omega_1$
and more repulsive above $\omega_1$ as shown in Fig. 2 of Ref.
\cite{Zeyher1} (Note
that the curve in that Figure includes a constant instantaneous 
contribution of -0.29). The Van Hove singularity is located near
$\delta =0.28$. There is nearly no effect of the van Hove singularity
on $T_c$ for the same reasons as in the above case of the incipient bond-order
wave instability. (The curves in Fig. 3 of Ref. \cite{Zeyher1}
were calculated without the
charge-charge term and, erroneously, without the factor 1/2 in the
instantaneous part. Correcting this error amounts essentially to lower
the curves in that Figure by about a factor 4).
Fig. 9 should be compared with Fig. 6. $T_c$ decreases
in both cases strongly with increasing $V$ and practically vanishes for
$V \geq J$. \par
We have also searched for superconducting instabilities with order
parameters which are odd in frequency. The sum over Matsubara frequencies 
is always zero in this case so that the constraint of having no
double occupancies of sites at the same time is automatically fulfilled
for the two particles of the Cooper pair. There exists no static
approximation for the kernel in this case. Possible instabilities are
again determined by the linearized gap equation Eq.(40) where $\Sigma_{an}$
and $\Theta$ have to be projected on the odd frequency parts. The transition
temperature is determined by the condition that the determinant of
a matrix consisting essentially of the kernel and the unity matrix is zero.
In Fig. 10 we have plotted this determinant as function of the temperature
for each of the five irreducible representations using $J=0.3$, $t'=0.$,
and $\delta =0.17$. The Figure clearly shows that none of the curves
tends to zero in the investigated temperature interval ruling out
any odd frequency pairing instability with a $T_c$ larger than $\sim 0.002$.
This also can be seen directly from Fig. 11 where the real part of
$\Theta_3(\omega + i \eta)$ is shown for odd frequency pairing with
$\Gamma_3$-symmetry corresponding also to triplet pairing. The effective
interaction is repulsive up to energies  $\sim t$ ruling out an 
instability towards superconductivity in this channel.

\subsection{Conclusions}
 
Sections II and III demonstrate the feasibility of developing 
a perturbation expansion for the $t-J$ model
in terms of X-operators and obtaining
explicit expressions for the leading contributions 
to the anomalous self-energy of physical electrons using an 1/N expansion.
The nonapplicability of Wick's theorem to X-operators does not cause
any problem in such an approach. One also should note that 
the more involved and sophisticated parts of sections II and III deal with 
contributions
which are far beyond those considered in previous treatments. For
instance, the slave boson 1/N result \cite{Grilli1}
 for the anomalous self-energy 
corresponds in our approach just to the inhomogenous term in the
integral equation Eq.(32). Similarly, only part of the first contribution
to the kernel, $\Theta^{(1)}$, was calculated in Ref. \cite{Onufrieva1}
using a diagram 
technique for X-operators. This shows in our opinion that the employed
functional approach is at least as suitable as other approaches to
treat highly correlated Fermi systems. 
Taking also the numerical results of
section 3 into account our main conclusions can be summarized 
as follows:\par

a) Our explicit expression for the O(1/N) anomalous self-energy is clearly
different from the corresponding expression of the slave boson
theory. In particular, the largest contribution to the retarded 
kernel comes in the present approach from the anomalous part of the 
vertex and has no analogue in
the slave boson approach. The presented expressions show for the
first time in an explicit way that the 1/N expansions are really 
different in the
two approaches. This difference can be traced back to different
Hilbert spaces and different enforcement of the constraint.
Even the order parameters for superconductivity are
not related in a simple way: The leading slave boson order parameter 
involves necessarily (small) violations of local constraints in order to be
nonzero whereas such violations are ruled out in our approach. \par
  
b) The kernel of the linearized gap equation consists of an instantaneous
and an retarded part and both are similar in magnitude at low 
frequencies and for momenta near the Fermi surface. We found that 
there are superconducting instabilities in each symmetry
channel and for all dopings.  The true ground state thus never describes 
a Fermi liquid but a superconductor similar as in the weak coupling 
case \cite{Kohn1}.
However, these instabilities are in general extremely weak leading 
to academically low transition temperatures. The only clear and  
robust exception is the d-wavelike $\Gamma_3$ symmetry where a strong
instability towards superconductivity occurs. Odd symmetry pairing
mechanisms turned out to be very weak and can be ruled out 
as a mechanism for high-T$_c$ superconductivity in our model.
\par

c) In the case of $\Gamma_3$ symmetry the real part of the retarded kernel
is weakly attractive at low frequencies
on an energy scale of $J$  or a fraction thereof and strongly
repulsive at higher frequencies whereas the instantaneous part is 
attractive. Solving numerically the linearized gap equation the
obtained transition temperatures $T_c$ are for $N=2$ of the order of
$0.01 |t|$ and thus in principle large enough to account for 
the phenomena of high-T$_c$ superconductivity. It is interesting to note 
that the Hubbard model at small or intermediate couplings also shows
d-wave superconductivity with similar values for $T_c$. \cite{Scalapino1} 
The instantaneous term is
instrumental in getting these large values for $T_c$: First, its cutoff is
given by the effective band width and thus in general larger than $J$.
Secondly, due to the large $T_c$, the solution of the gap equation
involves large frequencies where the retarded term is strongly repulsive. 
As a result the retarded term is of less importance because attractive 
and repulsive parts of it cancel each other to a large extent.
The dominance of the instantaneous part and the presence of two cutoffs 
lead to strong deviations from BCS-behavior. For instance, $T_c$
depends linearly and not exponentially on $J$ except at very small
values for $J$. \par

d) $T_c$ is rather insensitive to the addition of a second-nearest
neighbor hopping term $t'$ and to a Van Hove singularity. 
The latter
can be understood by noting that the solution of the gap equation
involves momentum and frequency averages
in the instantaneous and retarded part, respectively, so that 
singularities in the density of states are washed out. $T_c$ depends,
however, sensitively on a nearest-neighbor Coulomb repulsion $V$
and becomes very small if $V$ is substantially larger than $J$. \par

e) Our calculations are limited to dopings larger than $\delta_{BO}
(\sim 0.14$ for $J=0.3$)
where an instability towards an incommensurate  bond-order wave of d-symmetry
occurs. The associated soft mode causes $\lambda_3 \rightarrow -\infty$
for $\delta \rightarrow \delta_{BO}$ whereas $T_c$ is nearly unaffected.
In the underdoped regime $\delta < \delta_{BO}$ we expect a
competition between bond-order and antiferromagnetic fluctuations
which is beyond the leading order of the 1/N expansion considered in
this investigation.

{\bf Acknowledgement:} The second author (A. Greco) thanks the International
Bureau of the Federal Ministry for Education , Science, Research and
Technology for financial support (Scientific-technological cooperation between
Argentina and Germany, Project No. ARG AD 3P) and the MPI-FKF for 
hospitality. Both authors acknowledge useful discussions with P. Horsch.

\begin{figure}
\protect
\caption{
Dependence of various contributions to the total kernel $\Theta$ on the
second momentum $\bf k'$ along the Fermi line for a fixed first momentum
${\bf k}=(2.465,0.309)$; a) contributions $\Theta^{(1)},\Theta^{(2)}$;
b) contributions $\Theta^{(3)},\Theta^{(4)}$, and the total $\Theta$.}
\end{figure}

\begin{figure}
\protect
\caption{
Lowest eigenvalues $\lambda_i$ of the static kernel $\Theta$ for the 
five representations
$\Gamma_i$ of $C_{4v}$ as a function of the doping $\delta$.}
\end{figure}

\begin{figure}
\protect
\caption{
Negative imaginary part of the d-wave projected kernel $\Theta_3(\omega
+ i \eta)$ as a function of the frequency $\omega$ using $\eta=0.005$.}
\end{figure}


\begin{figure}
\protect
\caption{  
Real part of the d-wave projected kernel $\Theta_3(\omega+i \eta)$
as a function of the frequency $\omega$ for
$\eta=0.005$. Inset: The same for small frequencies
using $\eta=0.002$.}
\end{figure} 


\begin{figure}
\protect
\caption{  
Transition temperature $T_c$ (with $|t|$ as energy unit)
for $\Gamma_3$ pairing using the total kernel (squares), the 
instantaneous part (circles), the retarded part (dimaonds) of the kernel, 
and the total kernel without charge-charge term (triangles).}
\end{figure}


\begin{figure}
\protect
\caption{  
Transition temperature $T_c$ 
for $\Gamma_3$ pairing versus $V$ (circles) and versus $J$ (squares).}
\end{figure}


\begin{figure}
\protect
\caption{  
Transition temperature $T_c$ 
for $\Gamma_3$ pairing for $V=0.15$ using the total kernel (squares) and 
the instantaneous part of the kernel (circles).}
\end{figure}


\begin{figure}
\protect
\caption{  
Transition temperature $T_c$ 
for $\Gamma_3$ pairing for $J=0.2$ and $t'=-0.35$ using the total kernel 
(squares) and the 
instantaneous part of the kernel (circles).}
\end{figure}


\begin{figure}
\protect
\caption{  
Transition temperature $T_c$ for $J=0.2$, $t'=-0.35$, $\delta = 0.23$ as
a function of $V$ using the total kernel (squares) and the 
instantaneous part of the kernel (circles).}
\end{figure}


\begin{figure}
\protect
\caption{  
Determinant associated with the gap equation for odd frequency pairing
with symmetry $\Gamma_i$ as a function of temperature $T$.}
\end{figure} 


\begin{figure}
\protect
\caption{  
Real part of the odd frequency, $\Gamma_3$ kernel $\Theta_3(\omega+i \eta)$
as a function of the frequency $\omega$ for
$\eta=0.005$. }
\end{figure}

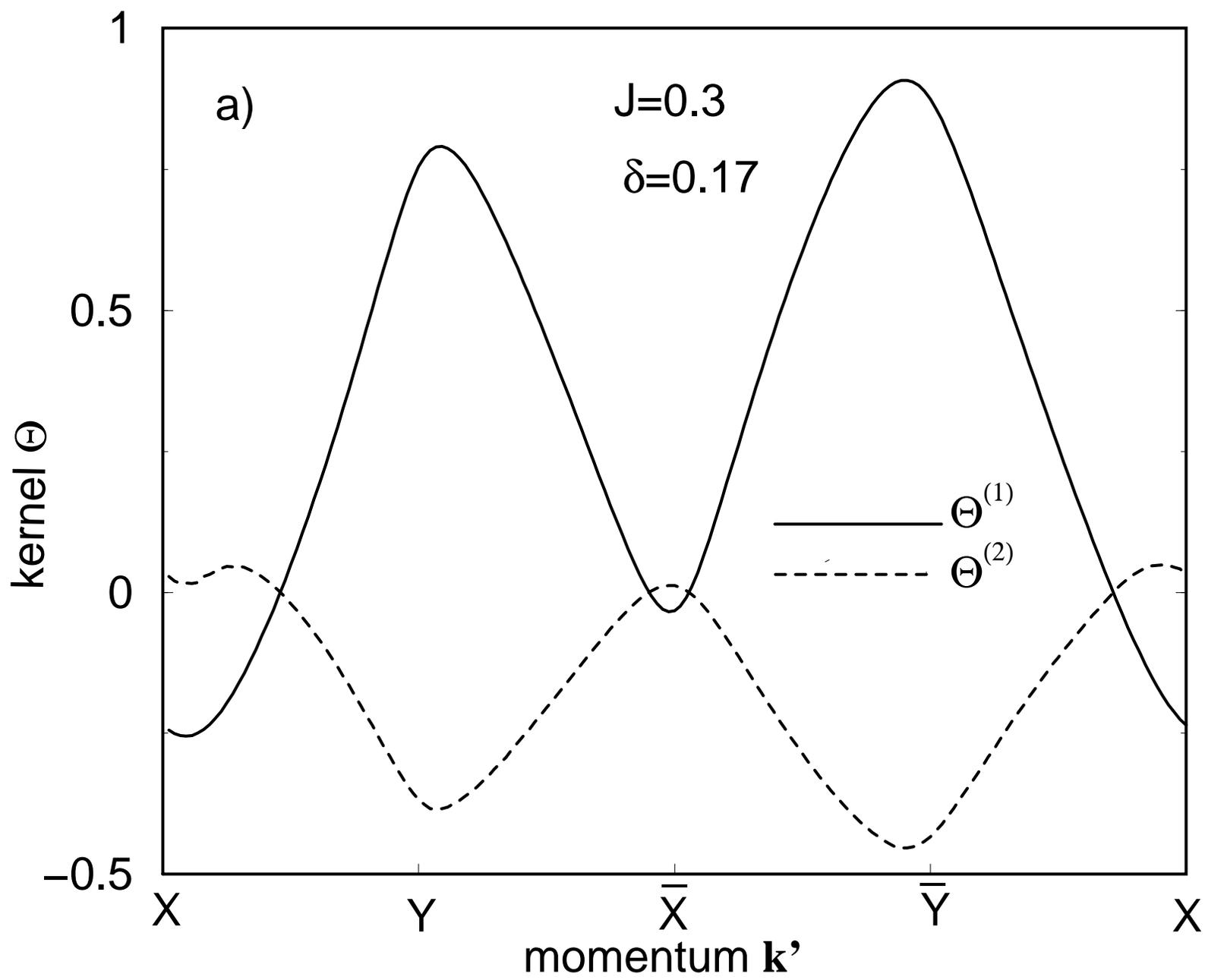
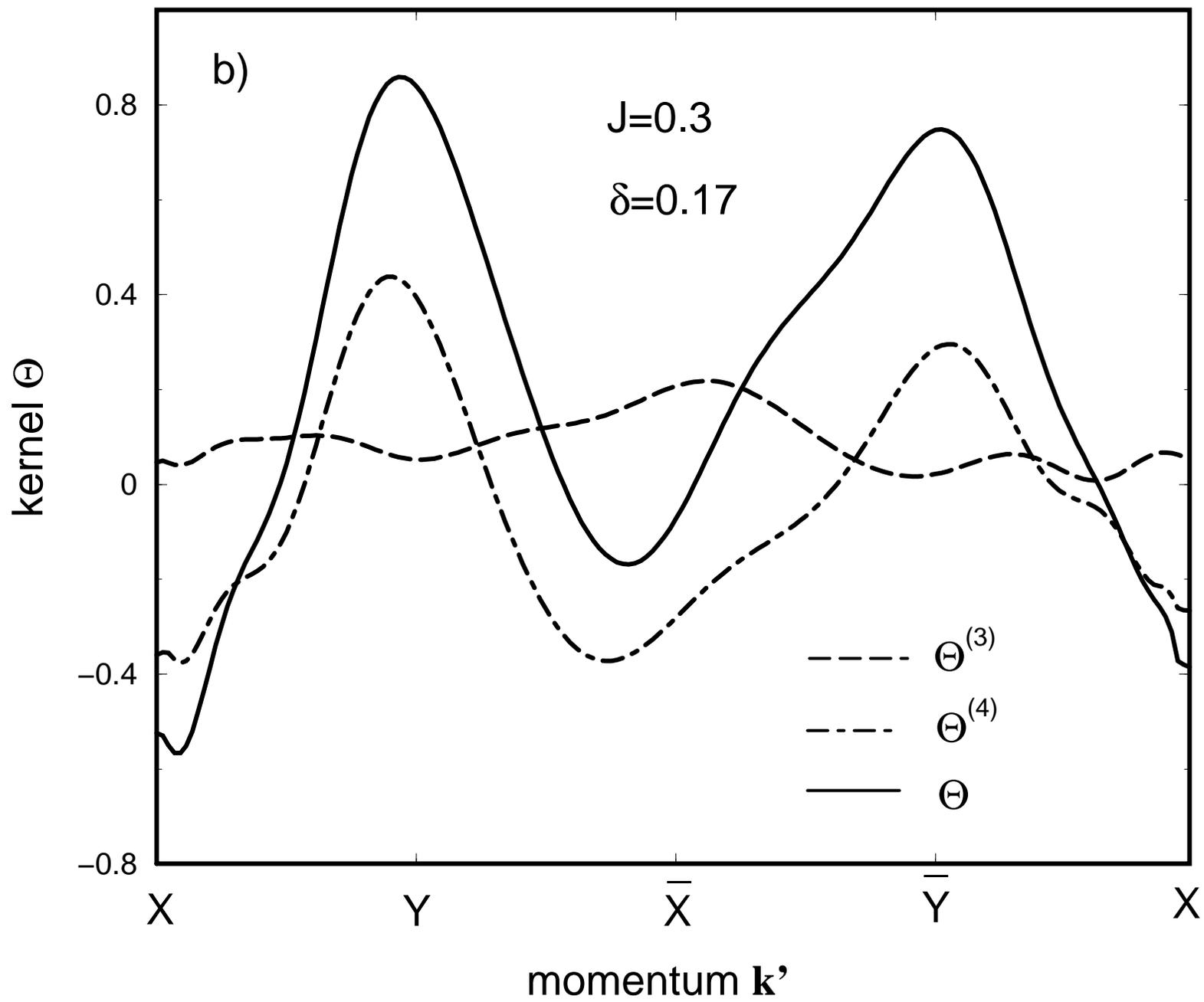
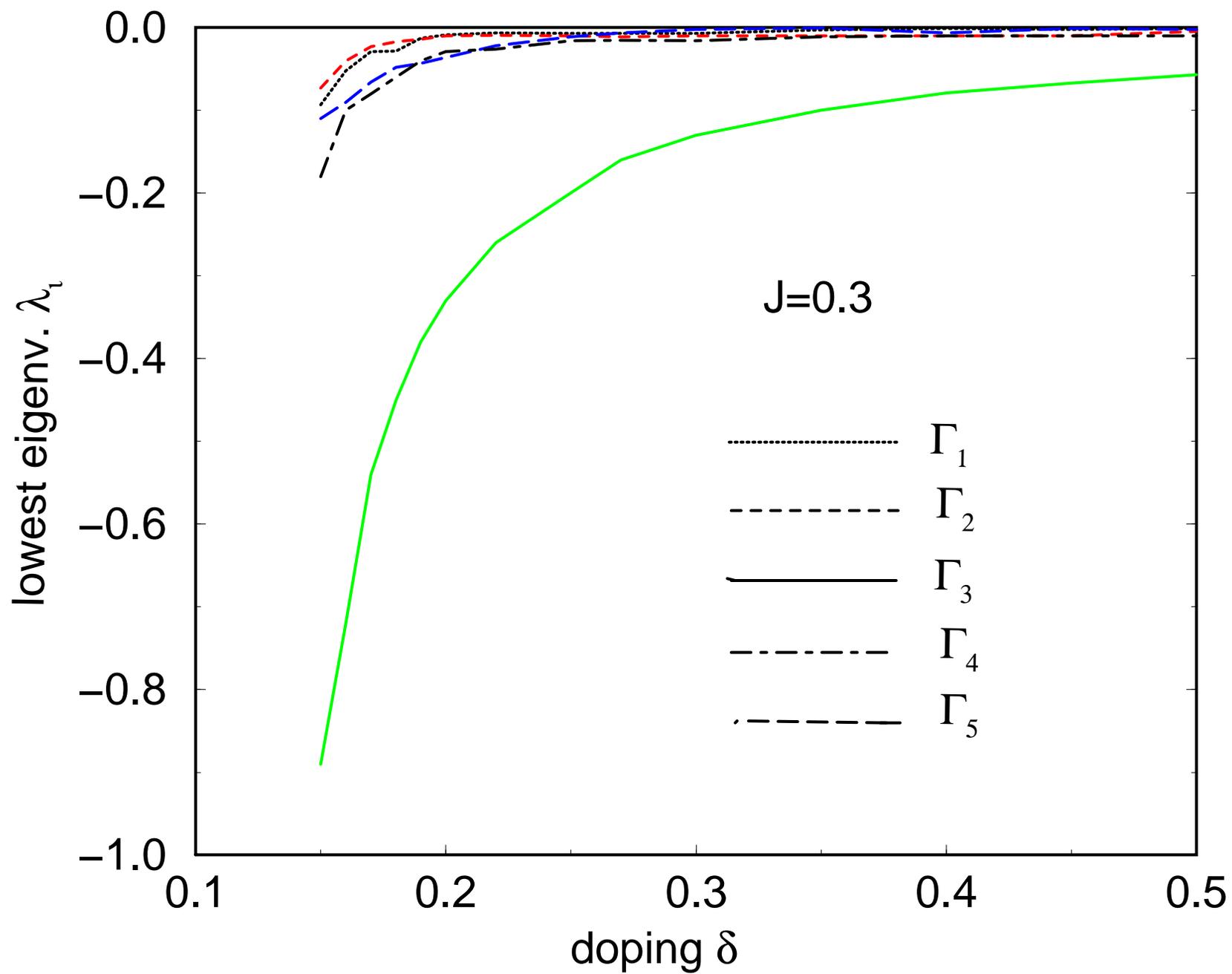
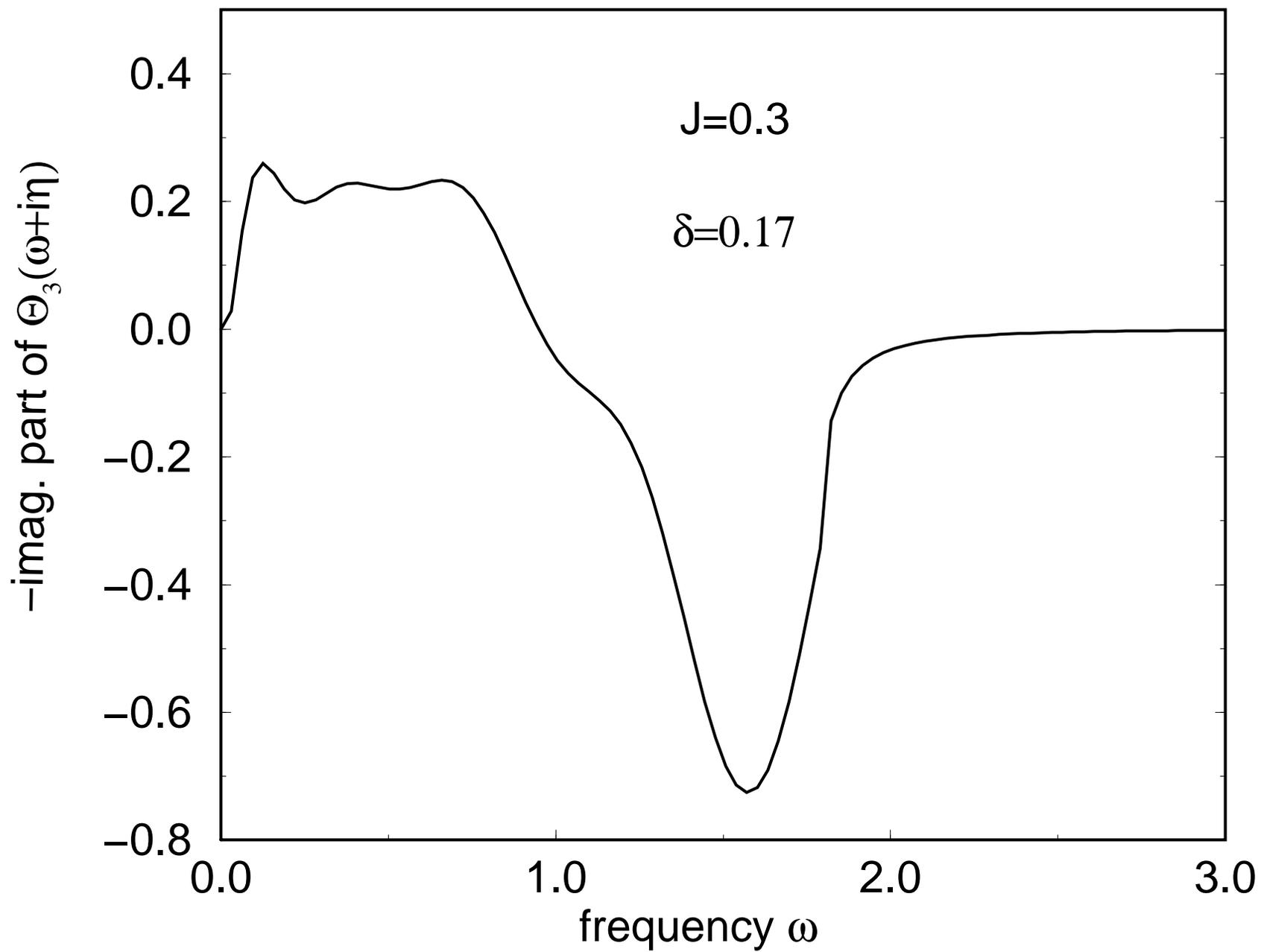
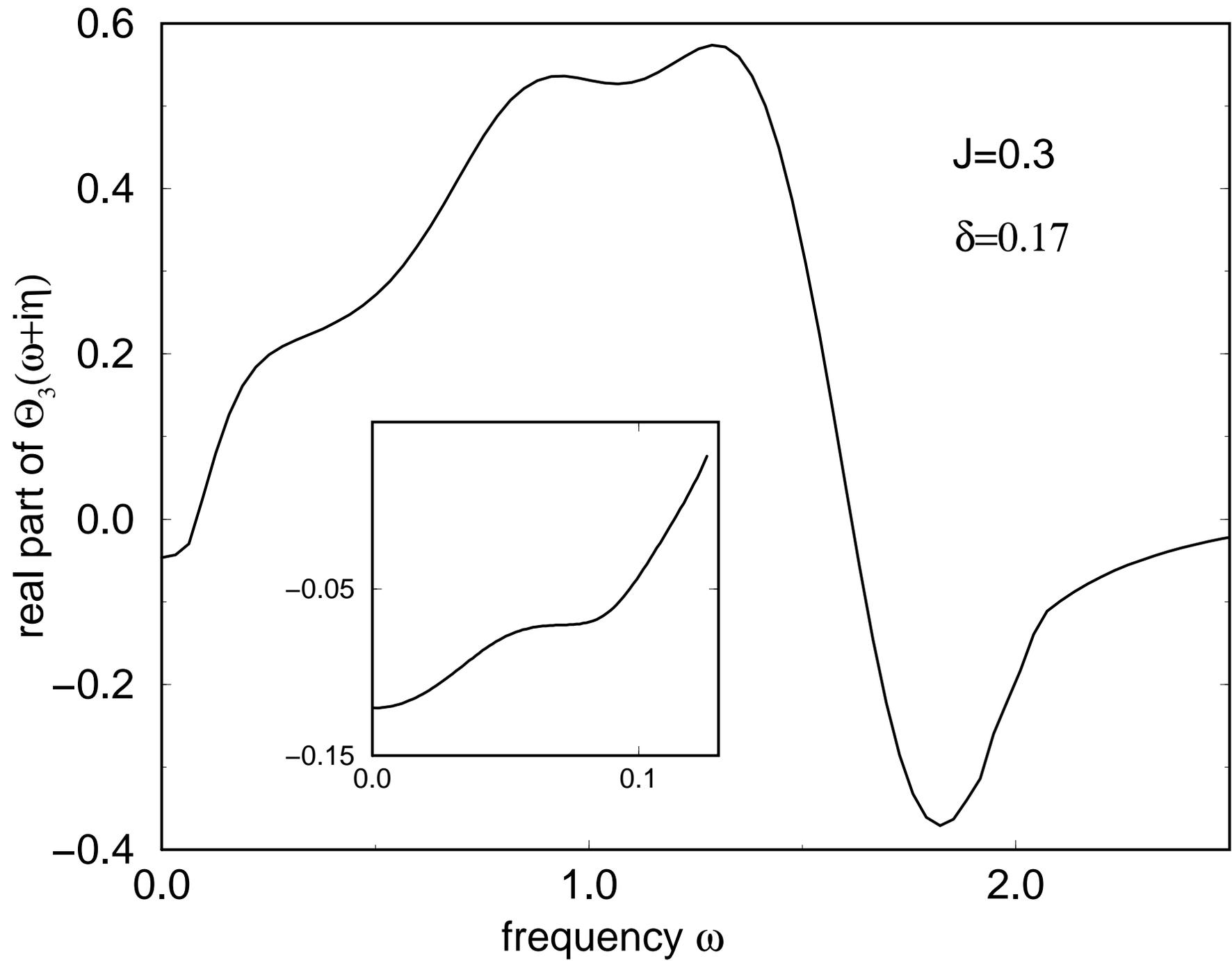
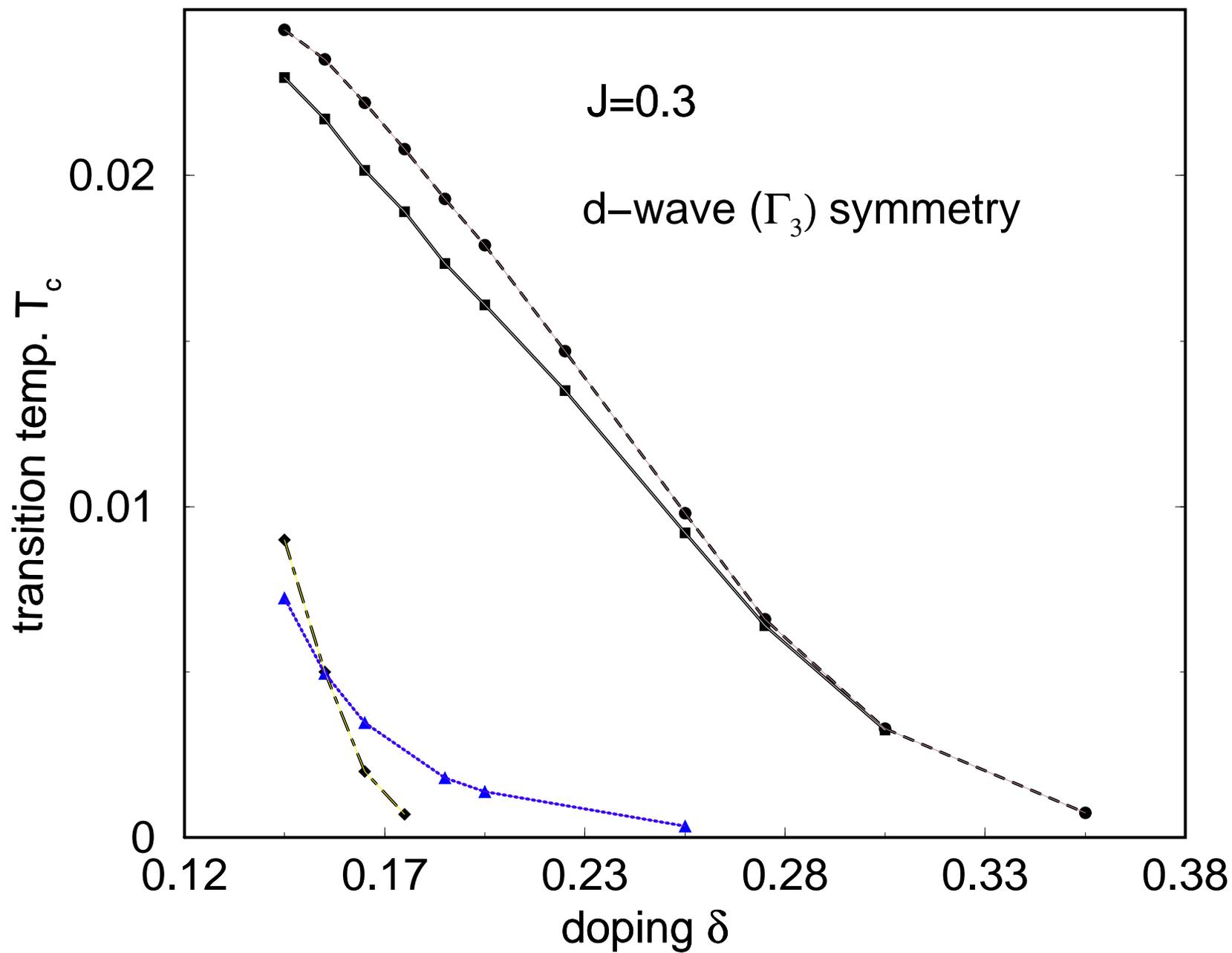
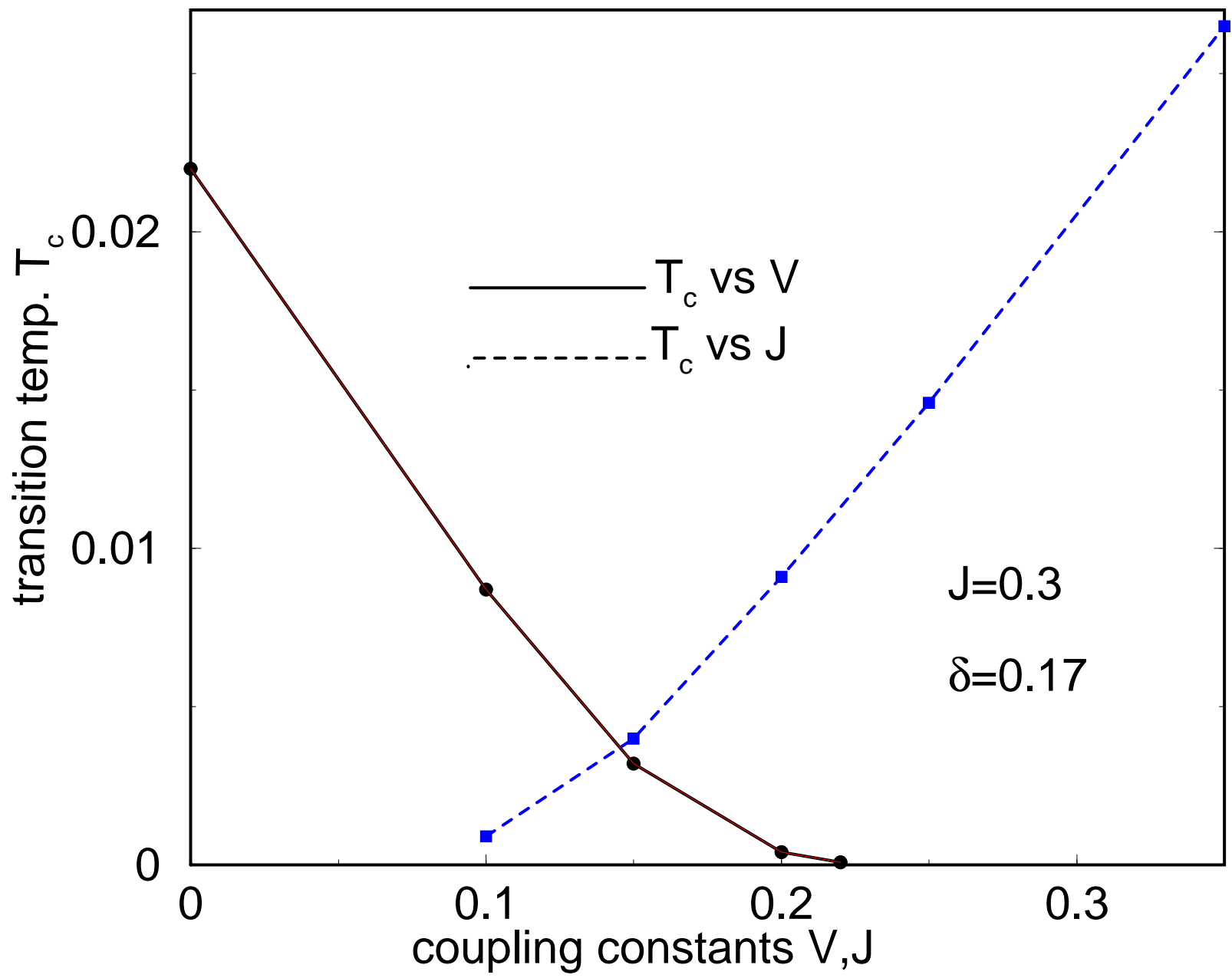
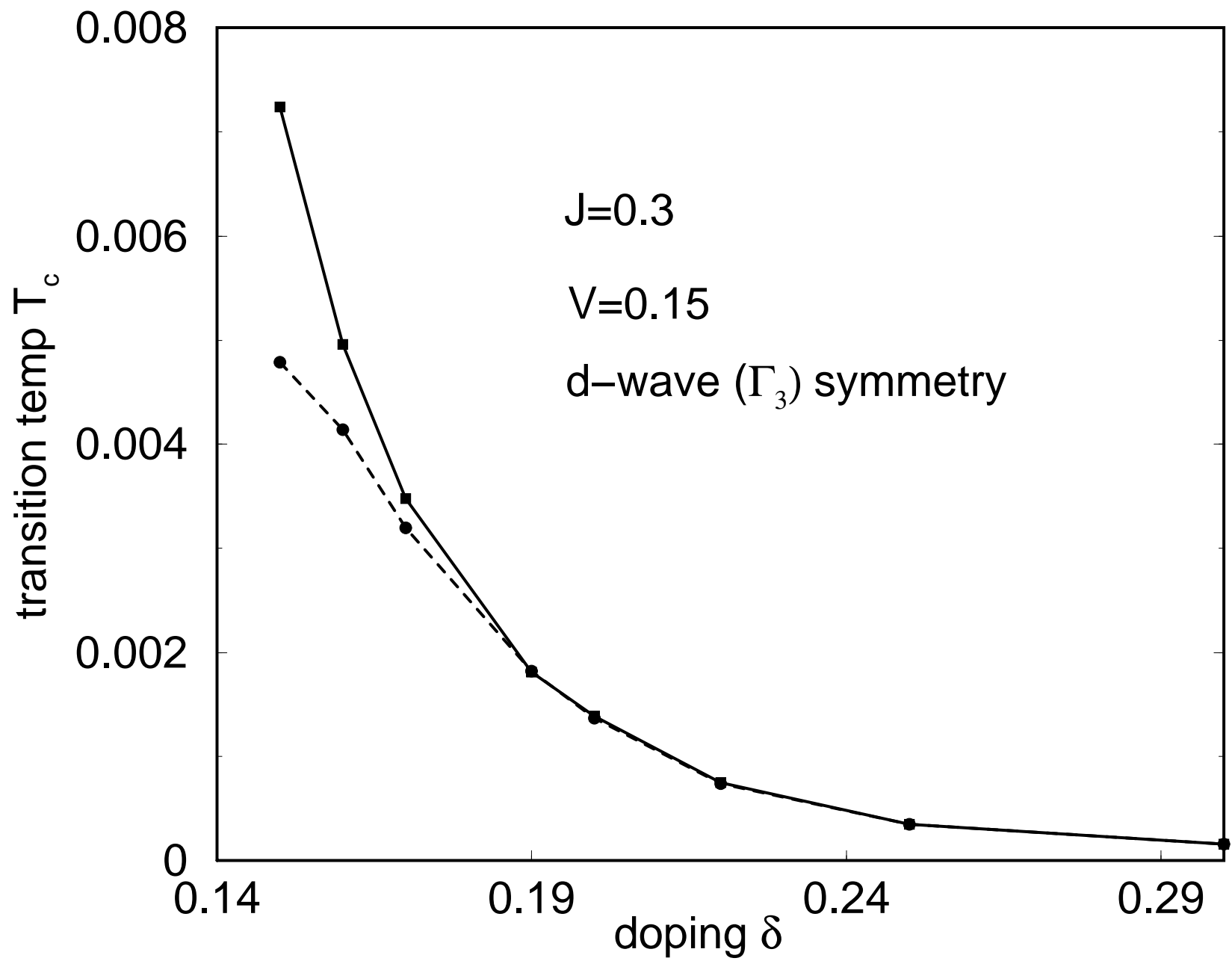
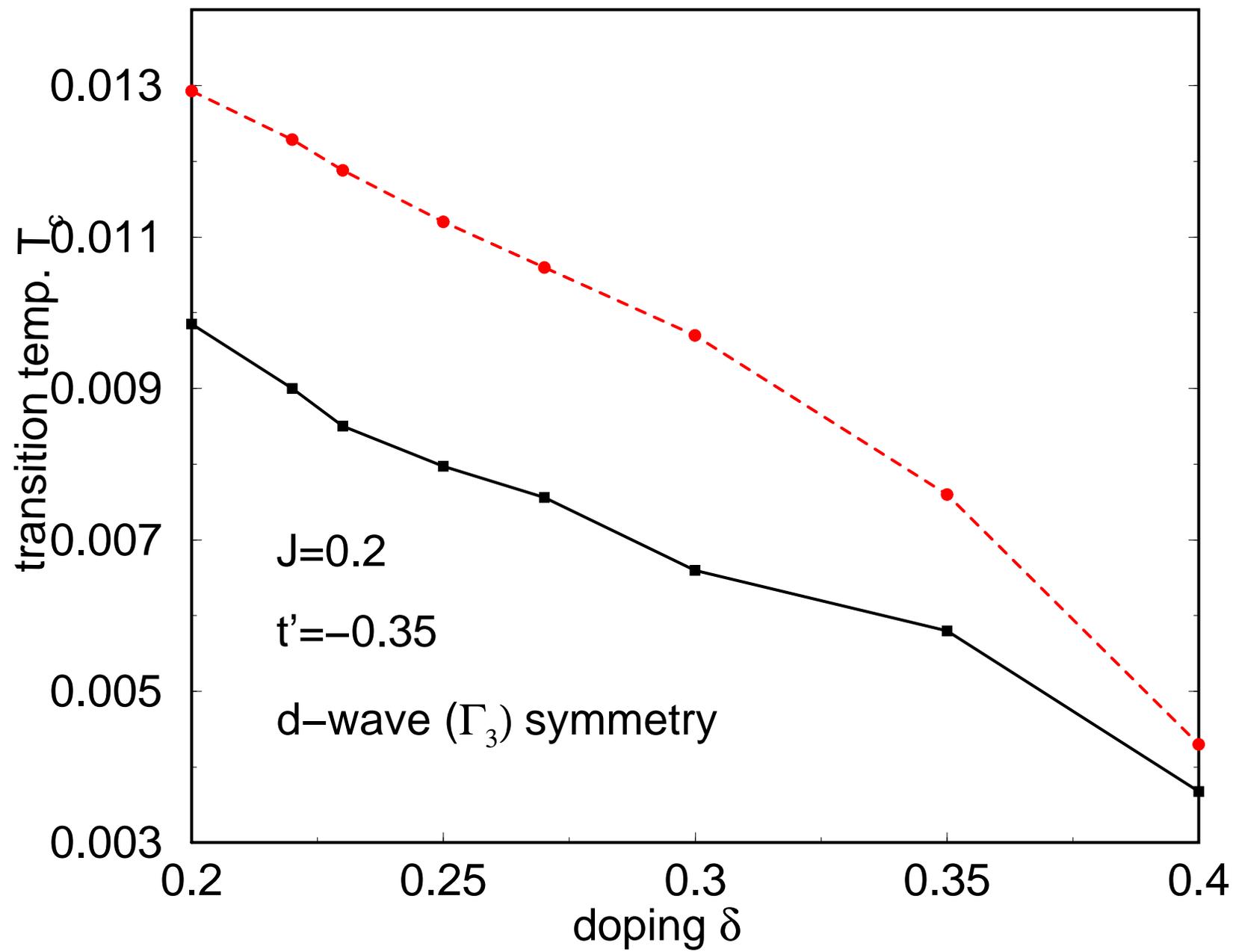
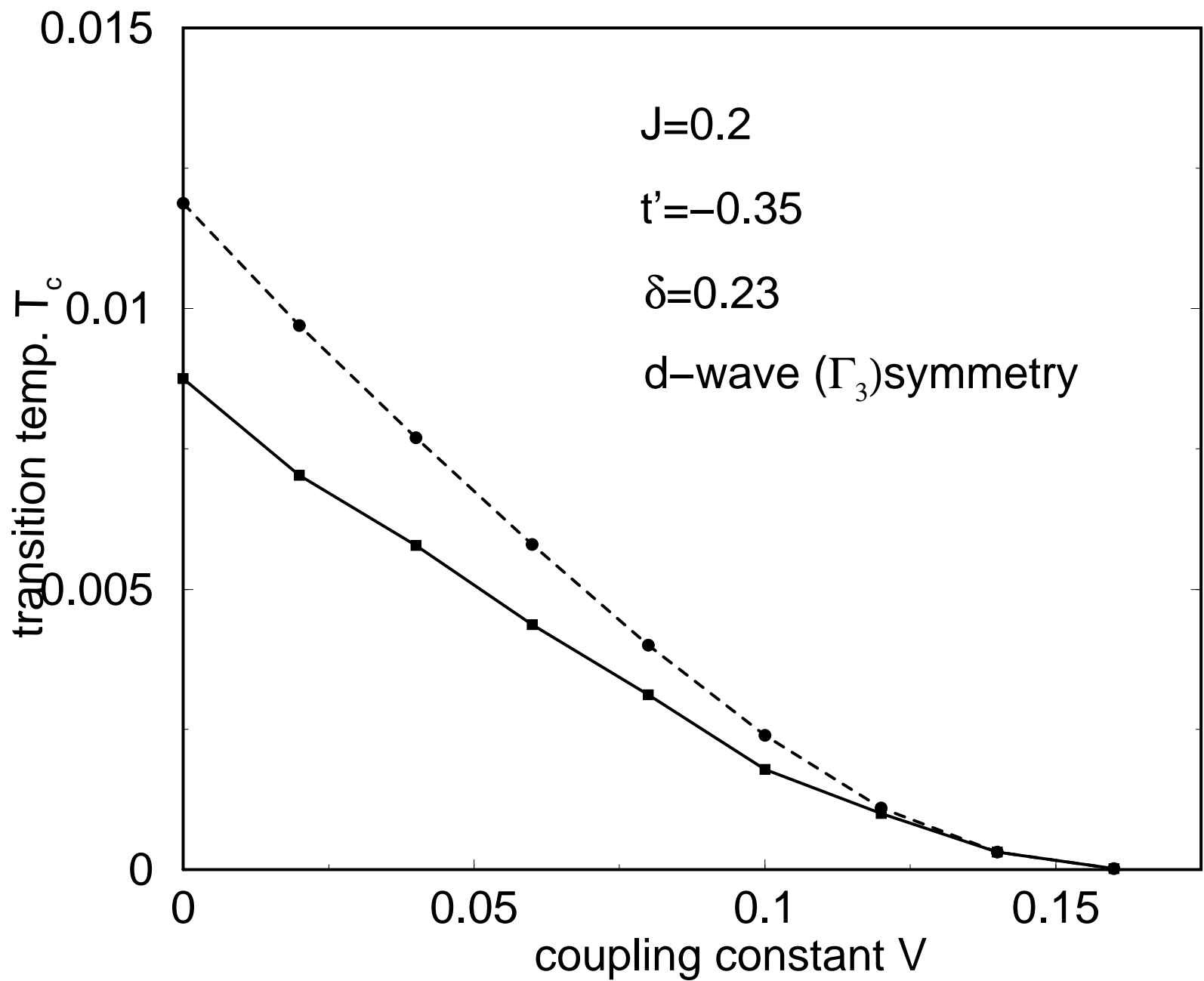
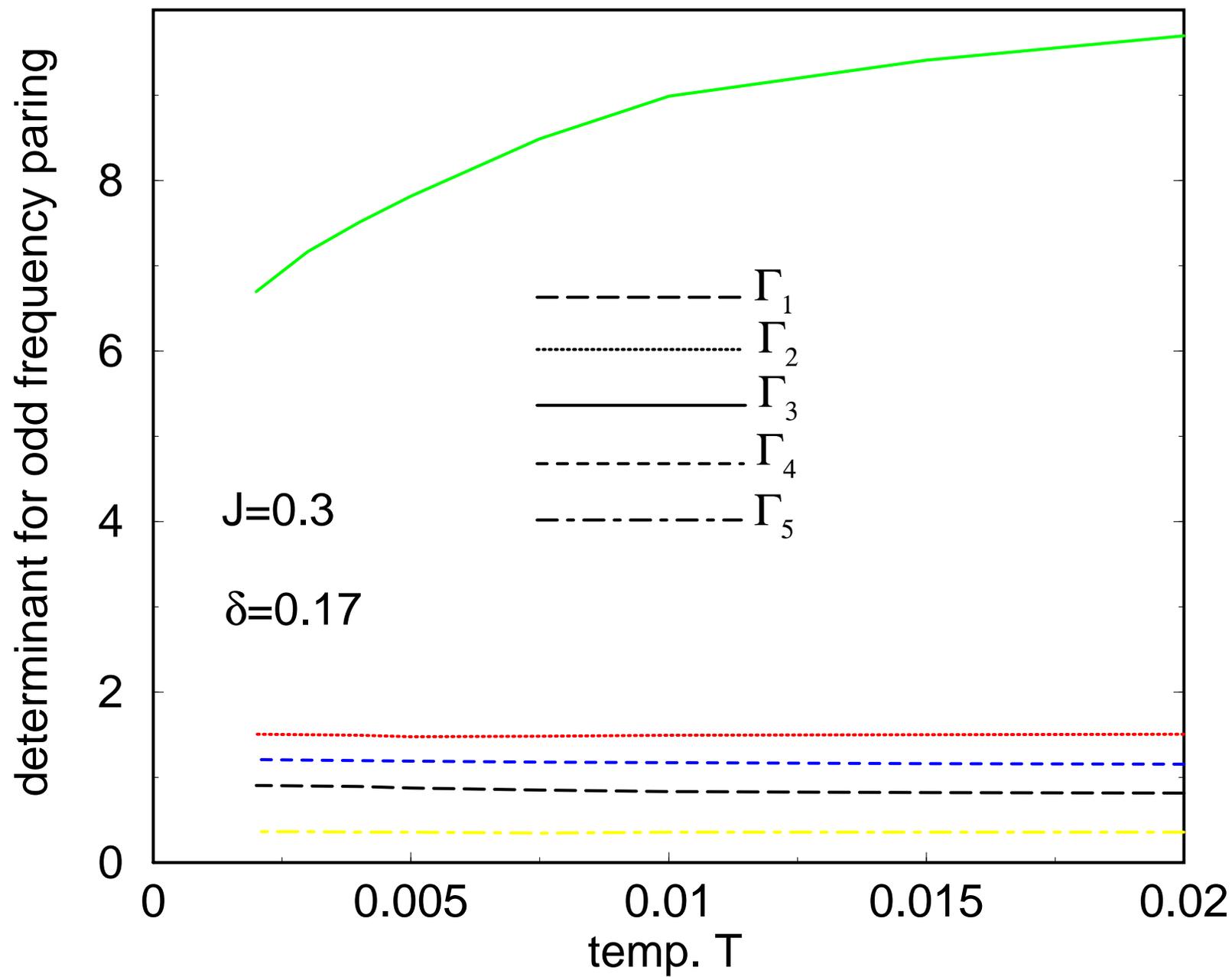
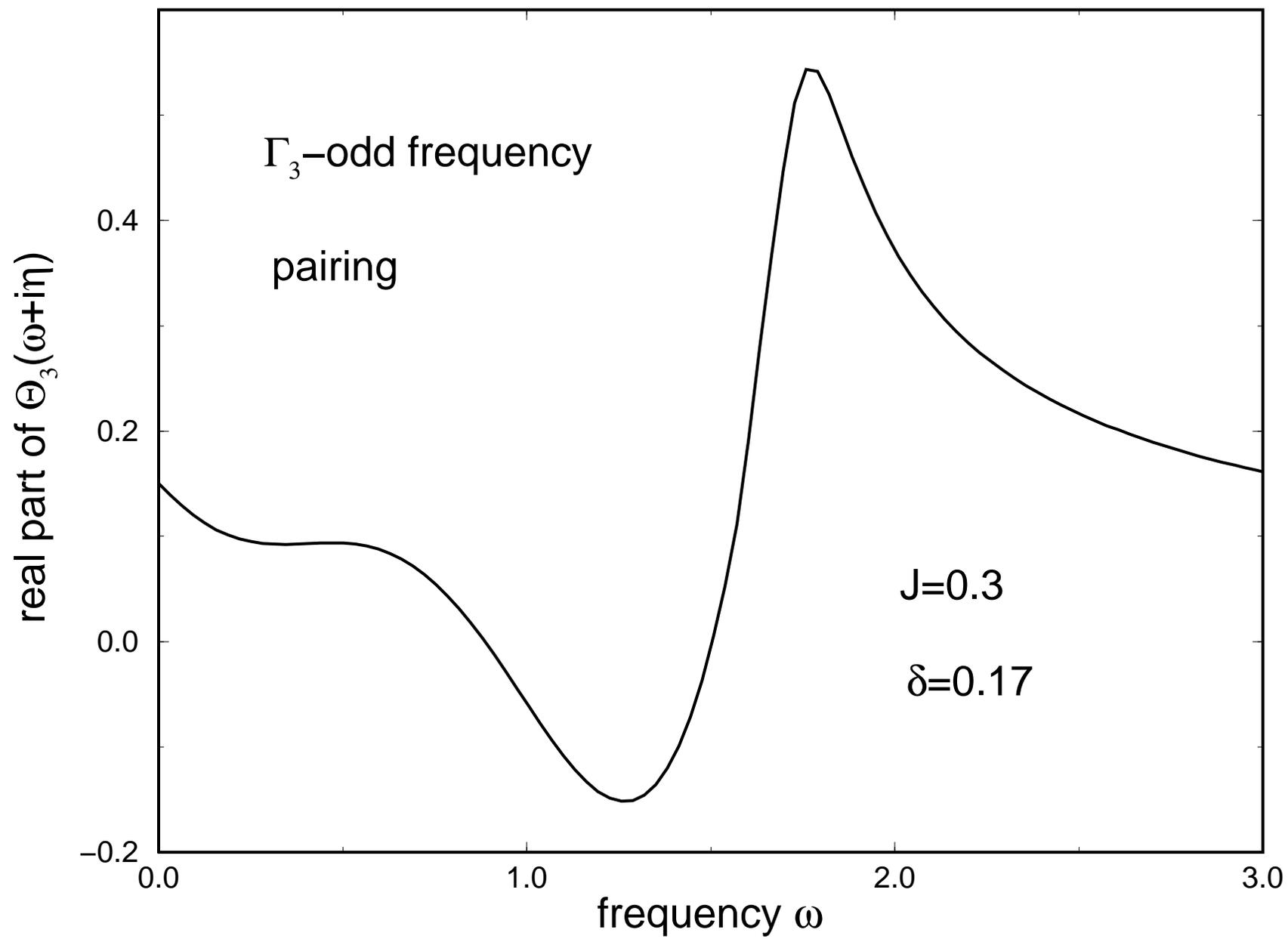
\end{document}